\documentclass{article}
\usepackage{spconf,amsmath,graphicx}
\usepackage{setspace}

\usepackage{url}
\usepackage{balance}
\usepackage{listings}
\usepackage{xcolor}
\usepackage{float}
\usepackage{graphicx}
\usepackage{braket}
\usepackage{amsmath}
\usepackage{amssymb}
\usepackage{array}
\usepackage{comment}
\usepackage{caption}
\usepackage{subcaption}
\usepackage{hyperref}
\usepackage{rotating}
\usepackage{cleveref}

\usepackage{pdflscape}

\definecolor{codegreen}{rgb}{0,0.6,0}
\definecolor{codegray}{rgb}{0.5,0.5,0.5}
\definecolor{codepurple}{rgb}{0.58,0,0.82}
\definecolor{backcolour}{rgb}{0.95,0.95,0.92}

\lstdefinestyle{mystyle}{
    commentstyle=\color{codegreen},
    keywordstyle=\color{magenta},
    numberstyle=\tiny\color{codegray},
    stringstyle=\color{codepurple},
    basicstyle=\ttfamily\footnotesize,
    breakatwhitespace=false,         
    breaklines=true,                 
    captionpos=t,                    
    keepspaces=true,                 
    numbers=left,                    
    numbersep=5pt,                  
    showspaces=false,                
    showstringspaces=false,
    showtabs=false,                  
    tabsize=2,
    frame=single,
    xleftmargin=7pt,
    xrightmargin=7pt
}

\graphicspath{{Images/}}

\title{Assembly to Quantum Compiler}

\address{}

\name{Andy Haverly$^{1}$, Shahram Rahimi$^{2}$, Mark A. Novotny$^{3}$}

\address{
    $^{1}$Mississippi State University, Department of Computer Science \& Engineering, arh876@msstate.edu \\
    $^{2}$The University of Alabama, Department of Computer Science, shahram.rahimi@ua.edu \\
    $^{3}$Mississippi State University, Department of Physics and Astronomy, man40@msstate.edu
}

\begin{document}
%
\maketitle

\begin{abstract}
This research presents a novel approach in quantum computing by transforming ARM assembly instructions for use in quantum algorithms. The core achievement is the development of a method to directly map the ARM assembly language, a staple in classical computing, to quantum computing paradigms. The practical application of this methodology is demonstrated through the computation of the Fibonacci sequence. This example serves to validate the approach and underscores its potential in simplifying quantum algorithms.
Grover’s Algorithm was realized through the use of quantum-specific instructions. These transformations were developed as part of an open-source assembly-to-quantum compiler (github.com/arhaverly/AssemblyToQuantumCompiler). This effort introduces a novel approach to utilizing classical instruction sets in quantum computing and offers insight into potential future developments in the field. The AssemblyToQuantumCompiler streamlines quantum programming and enables computer scientists to transition more easily from classical to quantum computer programming.

\end{abstract}
%
%


\section{Introduction}
\label{introduction}

Quantum computing extends beyond the limits of classical computation by exploiting fundamental principles such as superposition and entanglement. Superposition permits a quantum system to occupy multiple states concurrently until measurement, resulting in an exponential expansion of the computational state space as the number of qubits increases. Entanglement, a defining quantum property, allows pairs or groups of qubits to exist in correlated states such that the state of one qubit instantaneously influences the state of another, irrespective of spatial separation. This intrinsic correlation enables complex multi-qubit operations and forms the basis of quantum computing’s capacity to achieve exceptional computational efficiency. Through the exploitation of these phenomena, quantum computers can execute certain computations significantly faster and with fewer operations than classical systems, leading to substantial gains in computational capability and efficiency.

Many classical algorithms lack straightforward mappings to quantum algorithms due to the fundamental differences between classical and quantum computational models. Nevertheless, the ability to compile classical algorithms into quantum equivalents remains valuable. In the context of Grover’s algorithm, for instance, the oracle can be implemented using Boolean-like operations, as demonstrated in applications to the Boolean satisfiability problem \cite{grovers_original, grovers_sat}. Extending such Boolean-like operations broadens the range of problems to which Grover’s algorithm can be effectively applied.

Quantum gates are the basic components of quantum circuits, similar to logic gates in classical computing \cite{quantum_gates}. Many of these gates are listed below.

\subsection*{Pauli Gates}
\begin{itemize}
\item \textbf{X Gate}: Applies the Pauli-X transformation, corresponding to a $\pi$-radian rotation of the qubit about the X-axis.
\item \textbf{Y Gate}: Applies the Pauli-Y transformation, corresponding to a $\pi$-radian rotation of the qubit about the Y-axis.
\item \textbf{Z Gate}: Applies the Pauli-Z transformation, corresponding to a $\pi$-radian rotation of the qubit about the Z-axis.
\end{itemize}

\subsection*{Clifford Gates}
\begin{itemize}
\item \textbf{Hadamard (H) Gate}: Transforms a computational basis state into an equal superposition of $\ket{0}$ and $\ket{1}$.
\item \textbf{S Gate}: A phase-shift gate that applies a phase of $\pi/2$.
\item \textbf{Sdg Gate}: The Hermitian adjoint of the S gate.
\item \textbf{CNOT Gate}: Performs a controlled-NOT operation on a pair of qubits.
\item \textbf{SWAP Gate}: Exchanges the quantum states of two qubits.
\end{itemize}

\subsection*{Phase Gates}
\begin{itemize}
\item \textbf{T Gate}: Applies a phase shift of $\pi/4$.
\item \textbf{Tdg Gate}: The Hermitian adjoint of the T gate.
\item \textbf{U Gates}: General single-qubit unitary operations (U1, U2, U3) capable of representing any single-qubit transformation.
\end{itemize}

\subsection*{Rotation Gates}
\begin{itemize}
\item \textbf{RX Gate}: Performs a rotation of a qubit by a specified angle about the X-axis.
\item \textbf{RY Gate}: Performs a rotation of a qubit by a specified angle about the Y-axis.
\item \textbf{RZ Gate}: Performs a rotation of a qubit by a specified angle about the Z-axis.
\end{itemize}

\subsection*{Controlled Gates}
\begin{itemize}
\item \textbf{Controlled U Gates}: Including CU1, CU2, and CU3, which apply controlled versions of single-qubit unitary operations.
\item \textbf{CCX Gate (Toffoli Gate)}: A NOT operation controlled by two qubits.
\item \textbf{CSWAP Gate (Fredkin Gate)}: Executes a controlled swap between two qubits.
\end{itemize}

\subsection*{Advanced Quantum Gates}
\begin{itemize}
\item \textbf{Barrier}: A circuit directive used to delineate and prevent reordering of quantum operations.
\item \textbf{CRX, CRY, CRZ Gates}: Controlled single-axis rotation gates.
\item \textbf{Quantum Fourier Transform (QFT)–Related Gates}: Gates commonly employed in algorithms that rely on Fourier transforms.
\end{itemize}

\subsection*{Additional Quantum Gates in Qiskit \cite{qiskit}}
\begin{itemize}
\item \textbf{Multi-Controlled Gates}: Gates such as multi-controlled Toffoli (MCT) gates, which extend the CCX gate to an arbitrary number of control qubits.
\item \textbf{Ising Gates}: Including XX, YY, and ZZ gates, primarily used in quantum simulations of Ising-type Hamiltonians.
\item \textbf{RXX, RYY, RZZ Gates}: Two-qubit rotation gates about the XX, YY, and ZZ interaction axes.
\item \textbf{Reset Gate}: Reinitializes a qubit to the $\ket{0}$ state.
\item \textbf{Identity Gate (I)}: Implements the identity operation, leaving the qubit state unchanged.
\item \textbf{R Gate}: A general single-qubit rotation gate parameterized by both rotation angle and phase.
\item \textbf{SX Gate}: Represents the square root of the X gate, also referred to as the $\sqrt{\text{NOT}}$ gate.
\item \textbf{SXdg Gate}: The Hermitian adjoint of the SX gate.
\end{itemize}

These gates collectively form a foundational toolkit for exploiting quantum mechanical phenomena in computation, including superposition, entanglement, and quantum interference.

This manuscript presents a mapping from assembly instructions to quantum circuits, applies these circuits to compute the Fibonacci sequence, implements Grover’s algorithm using assembly-level instructions, introduces the open-source project associated with this work, and concludes with a summary of findings.

\section{Mapping Assembly Instructions to Quantum Circuits}
\label{mapping}

While many quantum gates are unique to quantum computation, several have clear analogues in classical assembly instructions. In this work, ARM Assembly Language is adopted as the classical reference \cite{arm_manual,arm_dev_guide}. Tables \ref{mapping1} and \ref{mapping2} present a correspondence between ARM assembly operations and their closest equivalents in quantum circuit form.

It should be noted that all subtraction operations discussed here employ reset operations to reduce overall circuit size. In scenarios where quantum coherence must be preserved, coherence can be maintained by introducing additional ancilla qubits and avoiding reset operations.

\begin{landscape}

\begin{table}[!htbp]
\centering
\caption{\label{TableMapping} Mapping of Assembly Instructions to Quantum Operations}
\label{mapping1}
\small
\begin{tabular}{|lp{3cm}p{3cm}l|p{4cm}lp{3cm}|}
\hline
\multicolumn{4}{|c|}{Classical} & \multicolumn{3}{c|}{Quantum} \\
\hline
Mnemonic & Instruction & Action & Example & ~Quantum Transformation & Coherence Maintaining? & Additional Notes \\
\hline
ADC & Add with carry & Rd := Rn + Op2 + Carry & ADC Rd, Rn, Operand2 & Figure \ref{fig:ADC_quantum} & Yes &  \\
ADD & Add & Rd := Rn + Op2 & ADD Rd, Rn, Rm & Figure \ref{fig:ADD_quantum} & Yes &  \\
AND & AND & Rd := Rn AND Op2 & AND Rd, Rn, Operand2 & Figure \ref{fig:AND_quantum} & Yes &  \\
B & Branch & R15 := address & B label & N/A & Yes & Unravel the branch \\
BEQ & Branch if Equal & R15 := address if Z flag is set & BEQ label & N/A & Yes & Unravel the branch, then use controlled gates to conditionally run operations \\
BIC & Bit Clear & Rd := Rn AND NOT Op2 & BIC Rd, Rn, Operand2 & Figure \ref{fig:BIC_quantum} & Yes &  \\
BL & Branch with Link & R14 := R15, R15 := address & BL label & N/A & N/A &  \\
BNE & Branch if Not Equal & R15 := address if Z flag is clear & BNE label & N/A & Yes & Unravel the branch, then use controlled gates to conditionally run operations \\
BX & Branch and Exchange & R15 := Rn, T bit := Rn[0] & Not translatable & N/A & N/A &  \\
CDP & Coprocessor Data Processing & (Coprocessor-specific) & (Coprocessor-specific) & N/A & N/A \\
CMN & Compare Negative & CPSR flags := Rn + Op2 & CMN Rn, Operand2 & Figure \ref{fig:CMN_quantum} & Yes &  \\
CMP & Compare & CPSR flags := Rn - Op2 & CMP Rn, Operand2 & Figure \ref{fig:CMP_quantum} & Yes &  \\
EOR & Exclusive OR & Rd := (Rn AND NOT Op2) OR (Op2 AND NOT Rn) & EOR Rd, Rn, Operand2 & Figure \ref{fig:EOR_quantum} & Yes &  \\
LDC & Load coprocessor from memory & Coprocessor load & (Coprocessor-specific) & N/A & N/A &  \\
LDM & Load multiple registers & Stack manipulation (Pop) & LDM Rn, {Rlist} & N/A & No &  \\
LDR & Load register from memory & Rd := (address) & LDR Rd, [Rn, Offset] & Figure \ref{fig:LDR_quantum} & No &  \\
LSL & Logical Shift Left & Rd := Rm LSL \#imm & LSL Rd, Rn, \#shift & Figure \ref{fig:LSL_quantum} & Yes &  \\
LSR & Logical Shift Right & Rd := Rm LSR \#imm & LSR Rd, Rn, \#shift & Figure \ref{fig:LSR_quantum} & Yes &  \\
MCR & Move CPU register to coprocessor register & cRn := rRn {<op>cRm} & (Coprocessor-specific) & N/A & N/A &  \\
MLA & Multiply Accumulate & Rd := (Rm * Rs) + Rn & MLA Rd, Rm, Rs, Rn & Figure \ref{fig:MLA_quantum} & Yes &  \\
\hline
\end{tabular}
\end{table}

\end{landscape}

\begin{landscape}

\begin{table}[!htbp]
\centering
\caption{\label{TableMapping} Mapping of Assembly Instructions to Quantum Operations Continued}
\label{mapping2}
\small
\begin{tabular}{|lp{3cm}p{3cm}l|p{4cm}lp{3cm}|}
\hline
\multicolumn{4}{|c|}{Classical} & \multicolumn{3}{c|}{Quantum} \\
\hline
Mnemonic & Instruction & Action & Example & ~Quantum Transformation & Coherence Maintaining? & Additional Notes \\
\hline
MOV & Move register or constant & Rd := Op2 & MOV Rd, Operand2 & Figure \ref{fig:MOV_quantum} & Yes &  \\
MRC & Move from coprocessor register to CPU register & Rn := cRn {<op>cRm} & (Coprocessor-specific) & N/A & N/A &  \\
MRS & Move PSR status/flags to register & Rn := PSR & MRS Rd, CPSR & Figure \ref{fig:MRS_quantum} & Yes &  \\
MSR & Move register to PSR status/flags & PSR := Rm & MSR CPSR, Rn & Figure \ref{fig:MSR_quantum} & Yes &  \\
MUL & Multiply & Rd := Rm * Rs & MUL Rd, Rm, Rs & Figure \ref{fig:MUL_quantum} & Yes &  \\
MVN & Move negative register & Rd := 0xFFFFFFFF EOR Op2 & MVN Rd, Operand2 & Figure \ref{fig:MVN_quantum} & Yes &  \\
ORR & OR & Rd := Rn OR Op2 & ORR Rd, Rn, Operand2 & Figure \ref{fig:ORR_quantum} & Yes &  \\
RSB & Reverse Subtract & Rd := Op2 - Rn & RSB Rd, Rn, Operand2 & Figure \ref{fig:RSB_quantum} & Yes &  \\
RSC & Reverse Subtract with Carry & Rd := Op2 - Rn - 1 + Carry & RSC Rd, Rn, Operand2 & Figure \ref{fig:RSC_quantum} & Yes &  \\
SBC & Subtract with Carry & Rd := Rn - Op2 - 1 + Carry & SBC Rd, Rn, Operand2 &  Figure \ref{fig:SBC_quantum} & Yes &  \\
STC & Store coprocessor register to memory & address := CRn & (Coprocessor-specific) & N/A & N/A &  \\
STM & Store Multiple & Stack manipulation (Push) & STM Rn, {Rlist} & N/A & No &  \\
STR & Store register to memory & <address> := Rd & STR Rd, [Rn, Offset] & Figure \ref{fig:STR_quantum} & No &  \\
SUB & Subtract & Rd := Rn - Op2 & SUB Rd, Rn, Operand2 & Figure \ref{fig:SUB_quantum} & Yes &  \\
SWI & Software Interrupt & OS call & SWI \#immediate & N/A & N/A &  \\
SWP & Swap register with memory & Rd := [Rn], [Rn] := Rm & SWP Rd, Rm, [Rn]  & N/A & No &  \\
TEQ & Test bitwise equality & CPSR flags := Rn EOR Op2 & TEQ Rn, Operand2 & Figure \ref{fig:TEQ_quantum} & Yes &  \\
TST & Test bits & CPSR flags := Rn AND Op2 & TST Rn, Operand2 & Figure \ref{fig:TST_quantum} & Yes &  \\
\hline
\end{tabular}
\end{table}

\end{landscape}

\begin{figure}[htp]
    \centering
    \includegraphics[width=8cm]{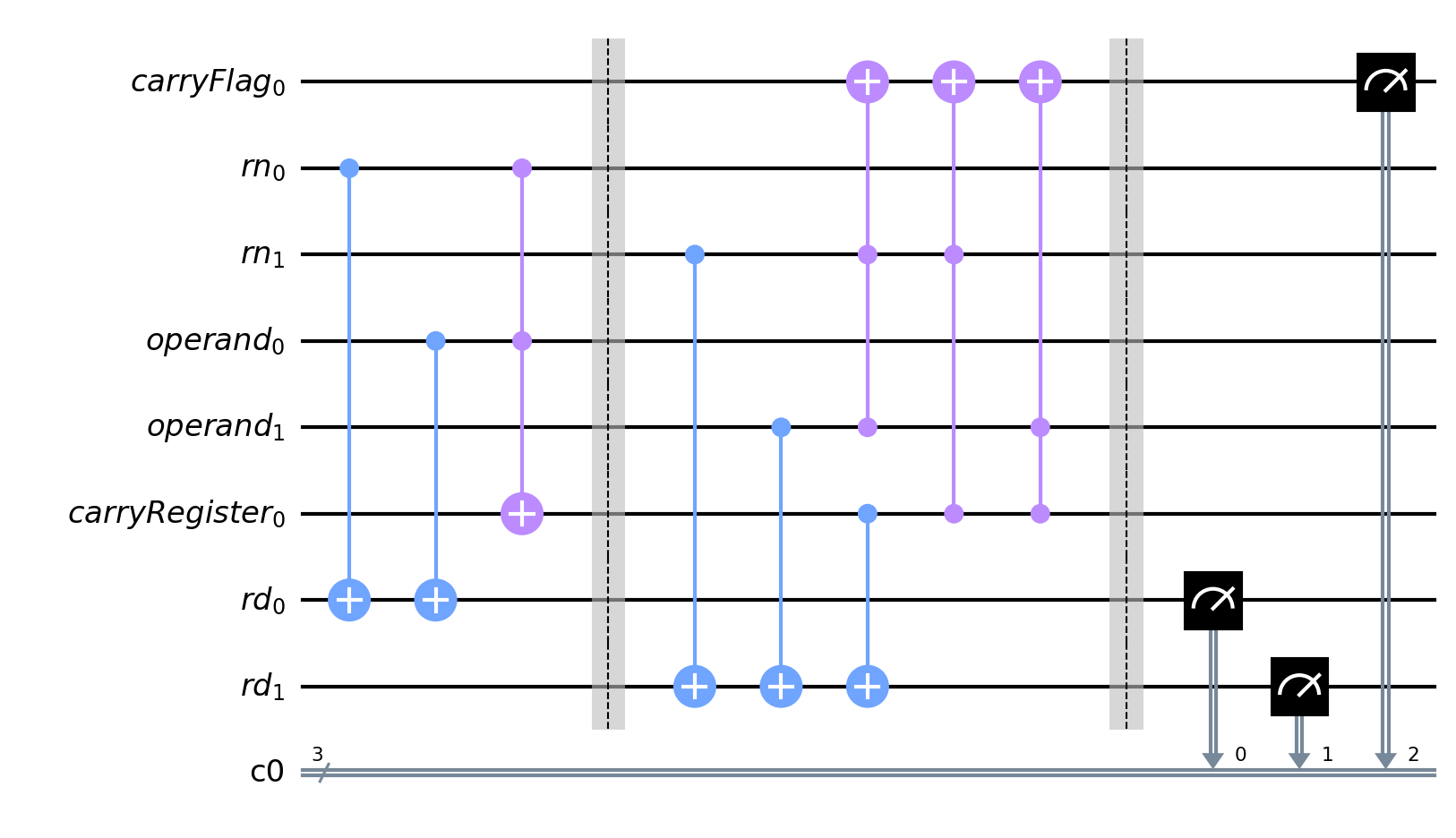}
    \caption{ADC Quantum Equivalent Circuit: A full or ripple carry adder that also sets the Carry Flag.}
    \label{fig:ADC_quantum}
\end{figure}

\textbf{ADC:} Figure \ref{fig:ADC_quantum} illustrates the mapping of the assembly-level ADC instruction to its quantum near-equivalent. The circuit consists of two input registers, \textit{rn} and \textit{operand}. A full or ripple-carry adder is employed to compute the sum, with the result stored in \textit{rd}. The \textit{carryFlag} is determined using an auxiliary \textit{carryRegister}.

\begin{figure}[htp]
    \centering
    \includegraphics[width=8cm]{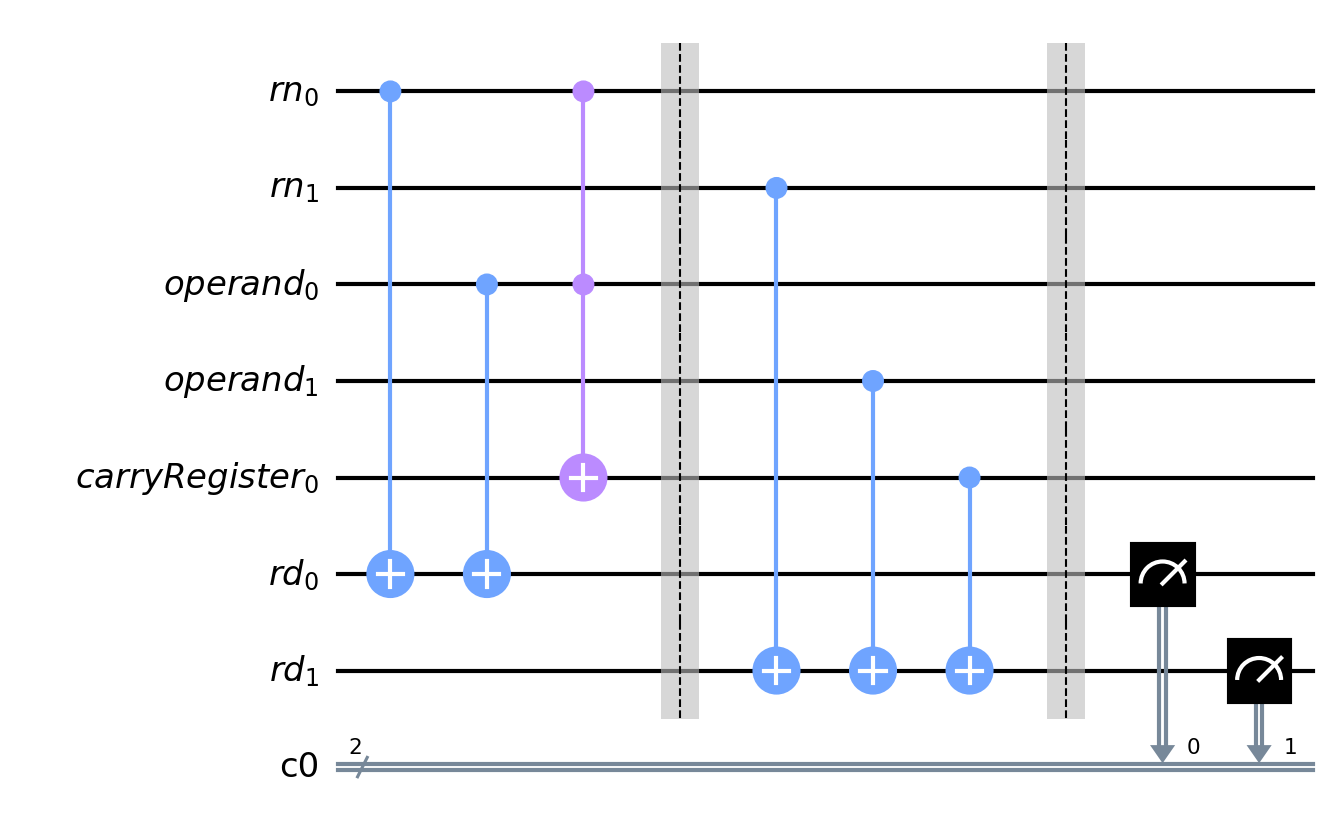}
    \caption{ADD Quantum Equivalent Circuit: A full or ripple carry adder.}
    \label{fig:ADD_quantum}
\end{figure}

\textbf{ADD:} Figure \ref{fig:ADD_quantum} depicts the mapping of the assembly ADD instruction to its quantum near-equivalent. The circuit utilizes two input registers, \textit{rn} and \textit{operand}, and employs a full or ripple-carry adder to compute the sum, which is stored in \textit{rd}.

\begin{figure}[htp]
    \centering
    \includegraphics[width=8cm]{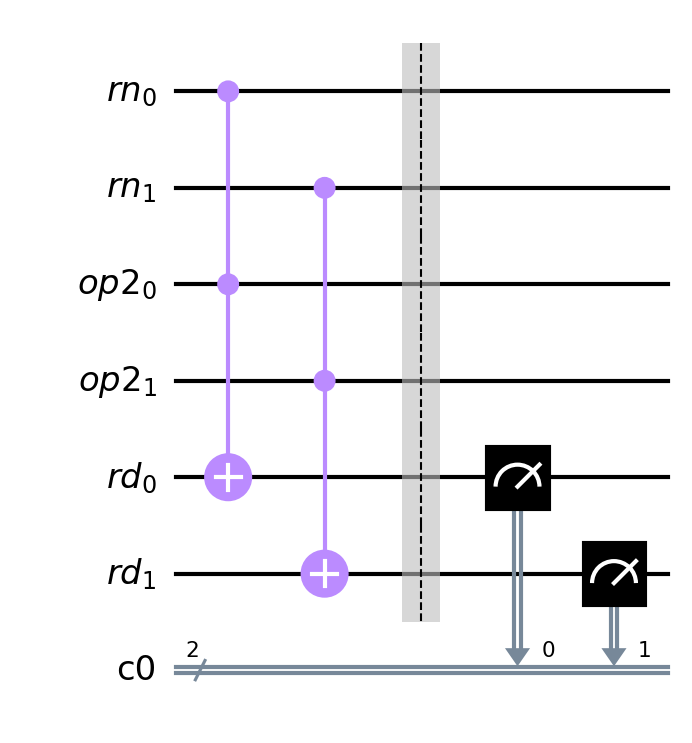}
    \caption{AND Quantum Equivalent Circuit: A Multiply-Controlled Toffoli gate is applied to each bit in the registers.}
    \label{fig:AND_quantum}
\end{figure}

\textbf{AND:} Figure \ref{fig:AND_quantum} illustrates the mapping of the assembly AND instruction to its quantum near-equivalent. The circuit features two input registers, \textit{rn} and \textit{op2}, and uses MCT gates to compute the AND operation, storing the result in \textit{rd}.

\begin{figure}[htp]
    \centering
    \includegraphics[width=8cm]{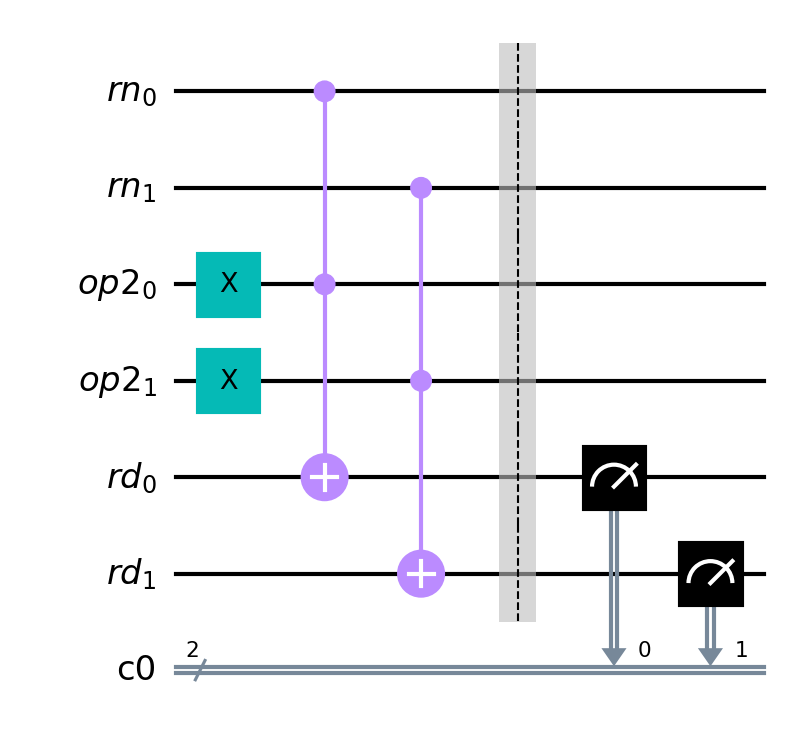}
    \caption{BIC Quantum Equivalent Circuit: Op2 is inverted then a Multiply-Controlled Toffoli gate is applied to each bit in the registers.}
    \label{fig:BIC_quantum}
\end{figure}

\textbf{BIC:} Figure \ref{fig:BIC_quantum} depicts the mapping of the assembly BIC instruction to its quantum near-equivalent. The circuit includes two input registers, \textit{rn} and \textit{op2}, and implements the BIC operation on \textit{rd} using a combination of X gates and MCT gates.

\begin{figure*}[htp]
    \centering
    \includegraphics[width=18cm]{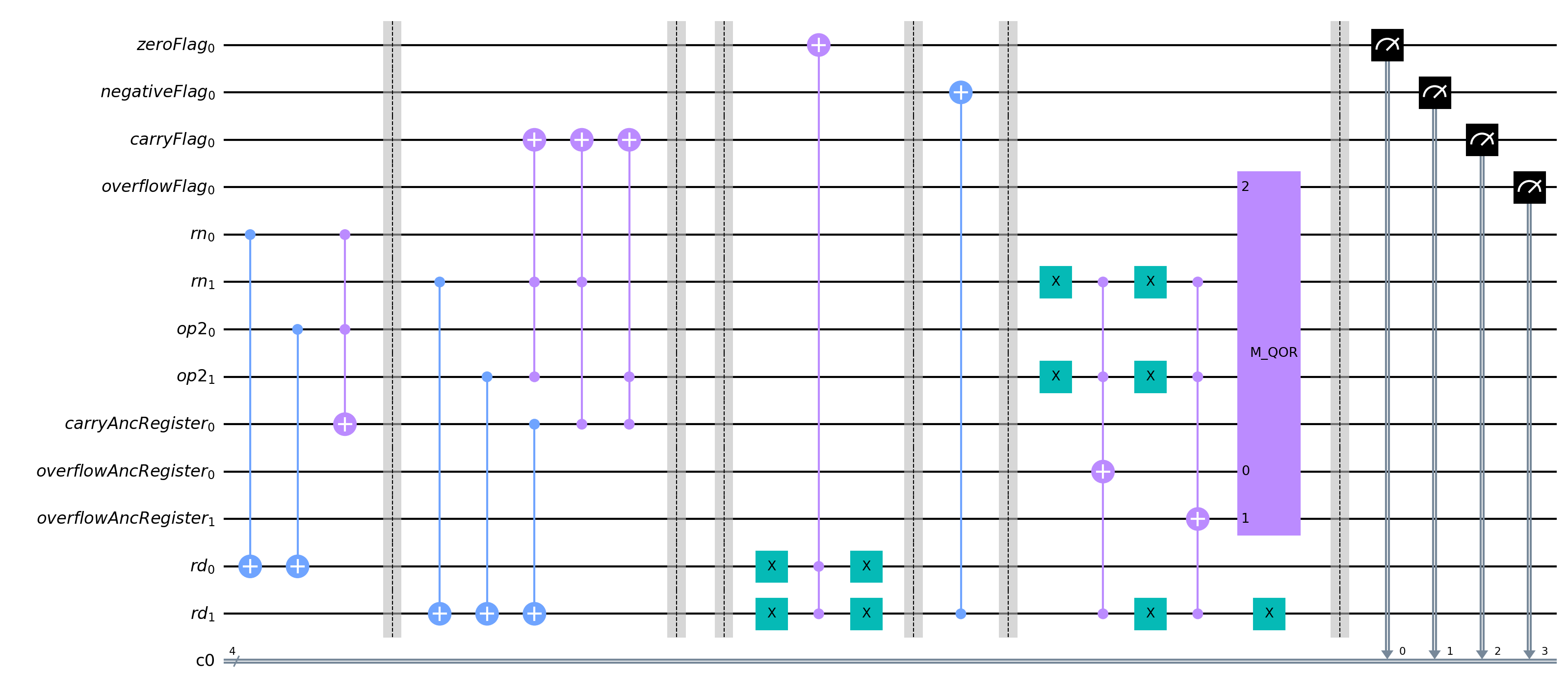}
    \caption{CMN Quantum Equivalent Circuit: Adds Rn and Op2 then determine the values for the Zero, Negative, Carry, and Overflow flags.}
    \label{fig:CMN_quantum}
\end{figure*}

\textbf{CMN:} Figure \ref{fig:CMN_quantum} illustrates the mapping of the assembly CMN instruction to its quantum near-equivalent. The circuit uses two input registers, \textit{rn} and \textit{op2}, and employs a full or ripple-carry adder to compute their sum. The \textit{Carry} flag is set if the addition produces a carry, the \textit{Zero} flag is determined by checking whether all result bits are zero, the \textit{Negative} flag is set based on the most significant bit (MSB), and the \textit{Overflow} flag is computed by XORing the MSBs of the input registers.

\begin{figure*}[htp]
    \centering
    \includegraphics[width=18cm]{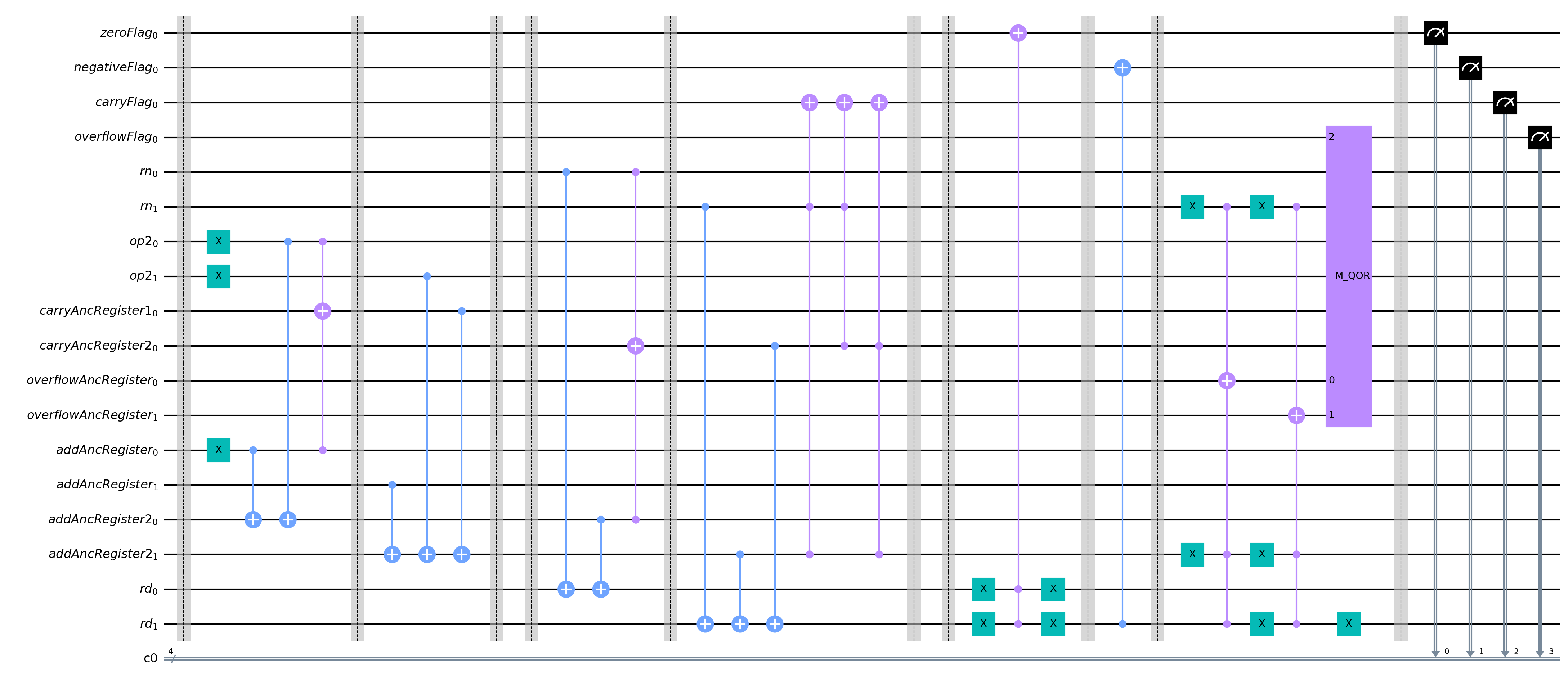}
    \caption{CMP Quantum Equivalent Circuit: Subtracts Op2 from Rn then determine the values for the Zero, Negative, Carry, and Overflow flags.}
    \label{fig:CMP_quantum}
\end{figure*}

\textbf{CMP:} Figure \ref{fig:CMP_quantum} depicts the mapping of the assembly CMP instruction to its quantum near-equivalent. The circuit consists of two input registers, \textit{rn} and \textit{op2}, and uses a full or ripple-carry adder to perform the subtraction of \textit{op2} from \textit{rn}. The \textit{Carry} flag is set if the operation generates a carry, the \textit{Zero} flag is determined by checking whether all result bits are zero, the \textit{Negative} flag is set based on the most significant bit (MSB), and the \textit{Overflow} flag is computed by XORing the MSBs of the input registers.

\begin{figure}[htp]
    \centering
    \includegraphics[width=8cm]{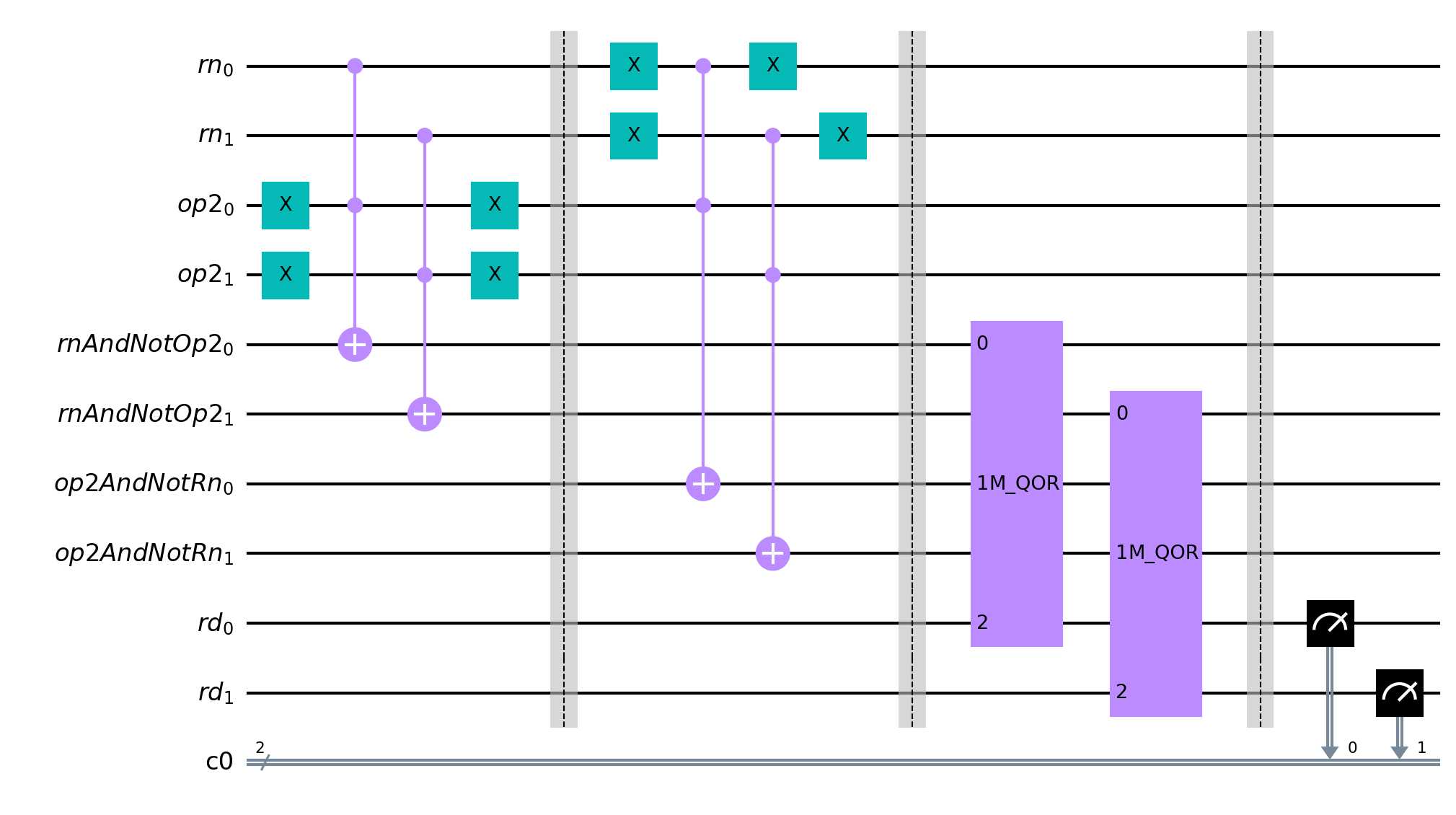}
    \caption{EOR Quantum Equivalent Circuit: The logic $(A \wedge \neg B) \vee (\neg A \wedge B)$ is implemented using X, Multiply-Controlled Toffoli, and Quantum OR gates.}
    \label{fig:EOR_quantum}
\end{figure}

\textbf{EOR:} Figure \ref{fig:EOR_quantum} illustrates the mapping of the assembly EOR instruction to its quantum near-equivalent. The circuit includes two input registers, \textit{rn} and \textit{op2}, and implements the bitwise XOR operation using a combination of X gates, MCT gates, and quantum OR gates.

\begin{figure}[htp]
    \centering
    \includegraphics[width=8cm]{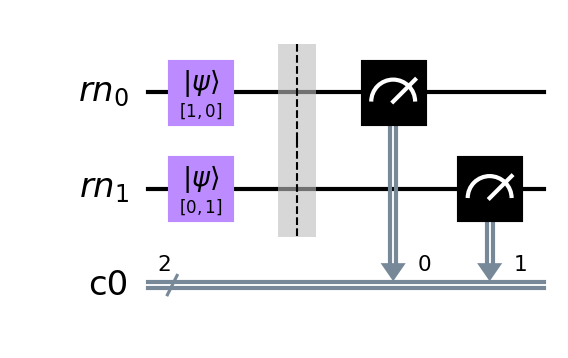}
    \caption{LDR Quantum Equivalent Circuit: Classically stores the value of the register.}
    \label{fig:LDR_quantum}
\end{figure}

\textbf{LDR:} Figure \ref{fig:LDR_quantum} depicts the mapping of the assembly LDR instruction to its quantum near-equivalent. The circuit uses a single input register, \textit{rn}, and employs an initialization gate to assign a classically known value to the register. If applied improperly, this operation can lead to decoherence.

\begin{figure}[htp]
    \centering
    \includegraphics[width=8cm]{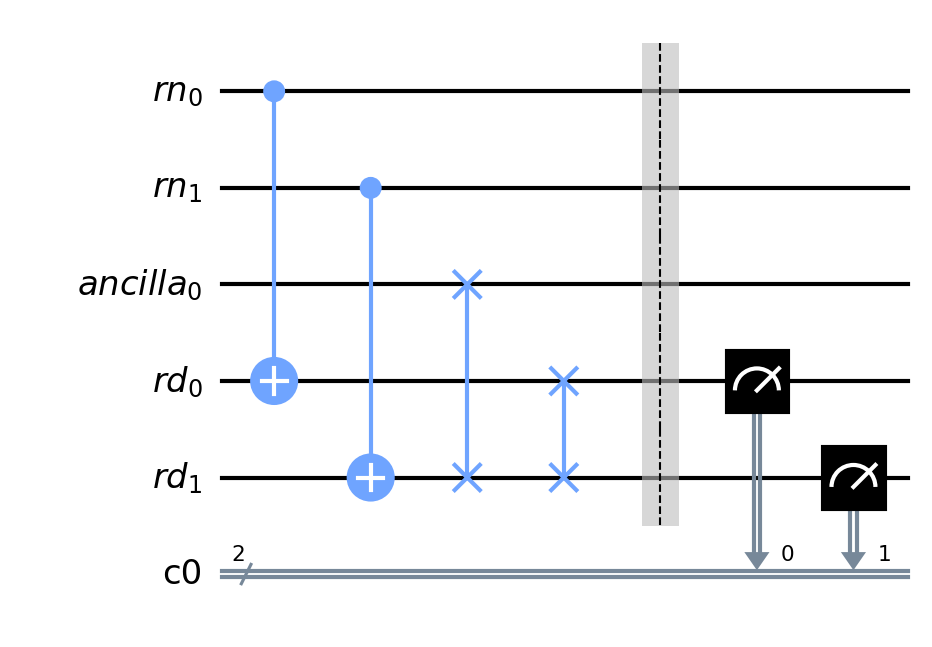}
    \caption{LSL Quantum Equivalent Circuit: Copies Rn to Rd then shifts using swap gates.}
    \label{fig:LSL_quantum}
\end{figure}

\textbf{LSL:} Figure \ref{fig:LSL_quantum} illustrates the mapping of the assembly LSL instruction to its quantum near-equivalent. The circuit utilizes a single input register, \textit{rn}, transfers its bits to \textit{rd}, and performs a left shift by a classically specified number of positions.

\begin{figure}[htp]
    \centering
    \includegraphics[width=8cm]{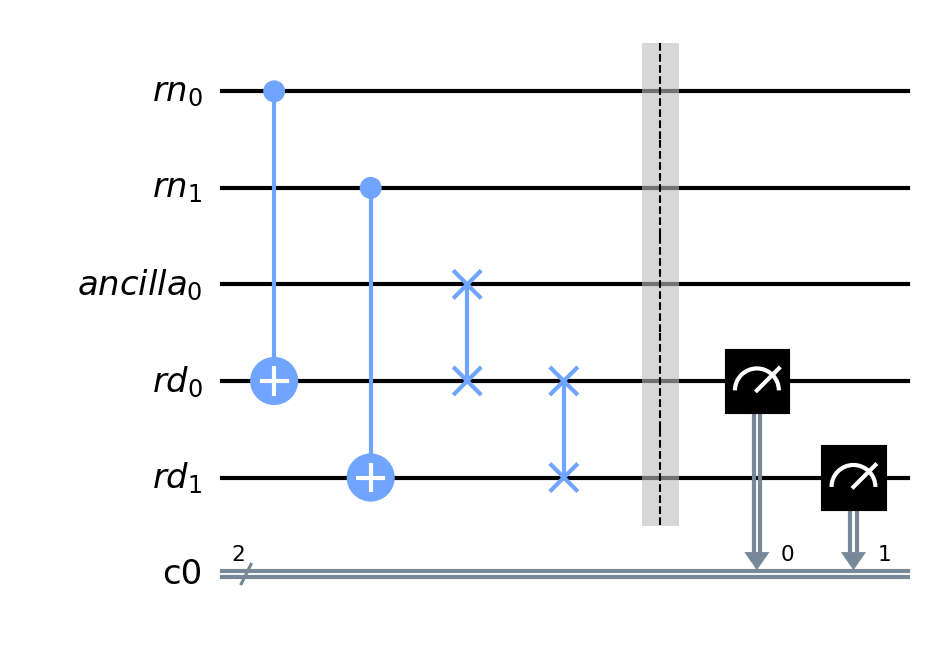}
    \caption{LSR Quantum Equivalent Circuit: Copies Rn to Rd then shifts using swap gates.}
    \label{fig:LSR_quantum}
\end{figure}

\textbf{LSR:} Figure \ref{fig:LSR_quantum} depicts the mapping of the assembly LSR instruction to its quantum near-equivalent. The circuit uses a single input register, \textit{rn}, transfers its bits to \textit{rd}, and performs a right shift by a classically specified number of positions.

\begin{figure*}[htp]
    \centering
    \includegraphics[width=18cm]{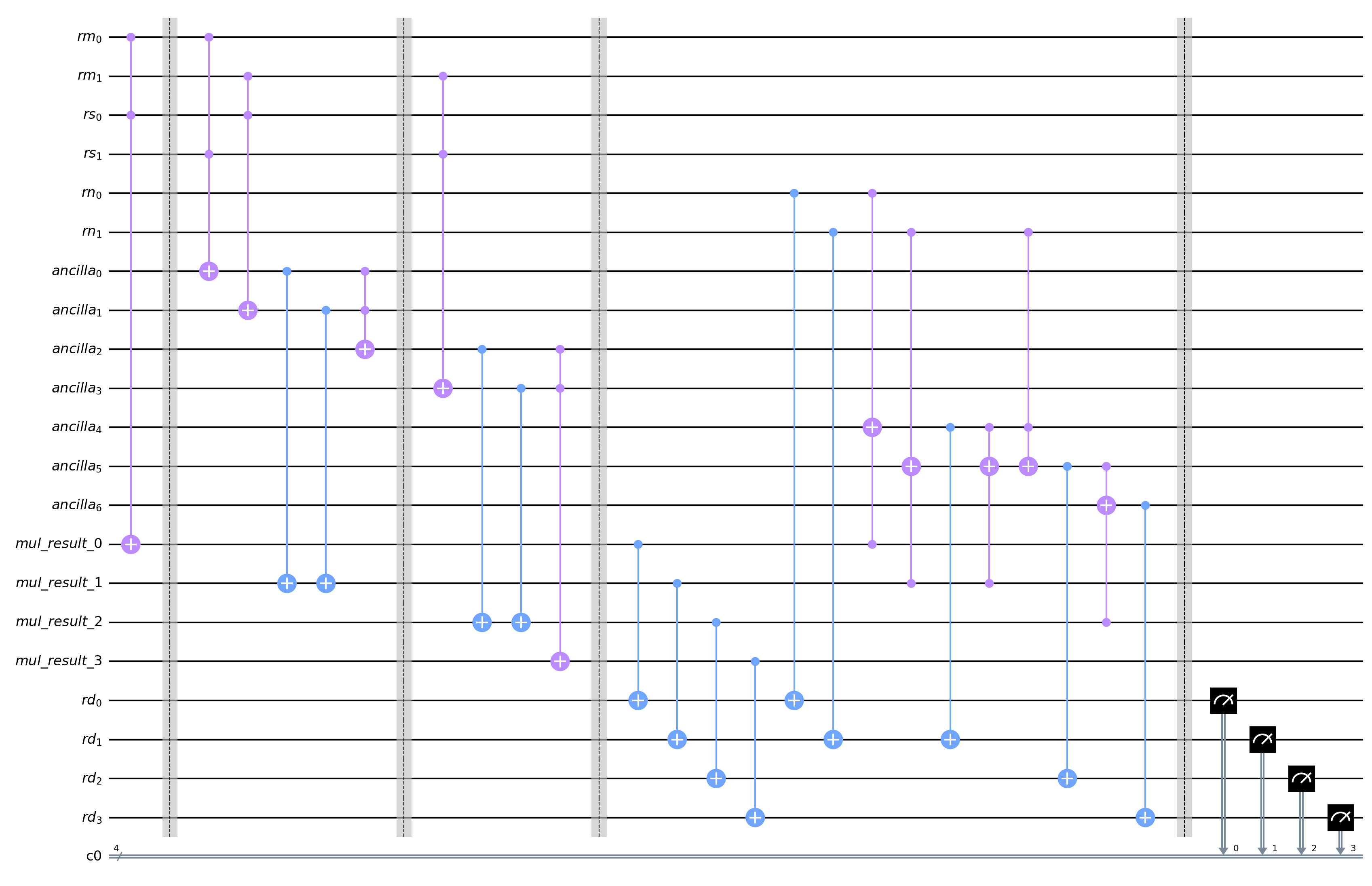}
    \caption{MLA Quantum Equivalent Circuit: The logic for multiplication is $R_0 \textrm{\&=} A_0 \land B_0, R_1 \textrm{\&=} (A_0 \land B_1) + (A_1 \land B_0), R_2 \textrm{\&=} (A_1 \land B_1) + \text{Carry from } R_1, R_3 \textrm{\&=} \text{Carry from } R_2$ then adds Rn using a ripple carry adder.}
    \label{fig:MLA_quantum}
\end{figure*}

\textbf{MLA:} Figure \ref{fig:MLA_quantum} illustrates the mapping of the assembly MLA instruction to its quantum near-equivalent. The circuit employs three input registers, \textit{rm}, \textit{rs}, and \textit{rn}. Boolean operations are used to compute each bit of the multiplication efficiently according to the logic $R_0 \textrm{\&=} A_0 \land B_0, R_1 \textrm{\&=} (A_0 \land B_1) + (A_1 \land B_0), R_2 \textrm{\&=} (A_1 \land B_1) + \text{Carry from } R_1, R_3 \textrm{\&=} \text{Carry from } R_2$ implemented with CNOT and MCT gates. Following the multiplication, a ripple-carry adder is used to add the value of \textit{rn} to the result in \textit{rd}.

\begin{figure*}[htp]
    \centering
    \includegraphics[width=18cm]{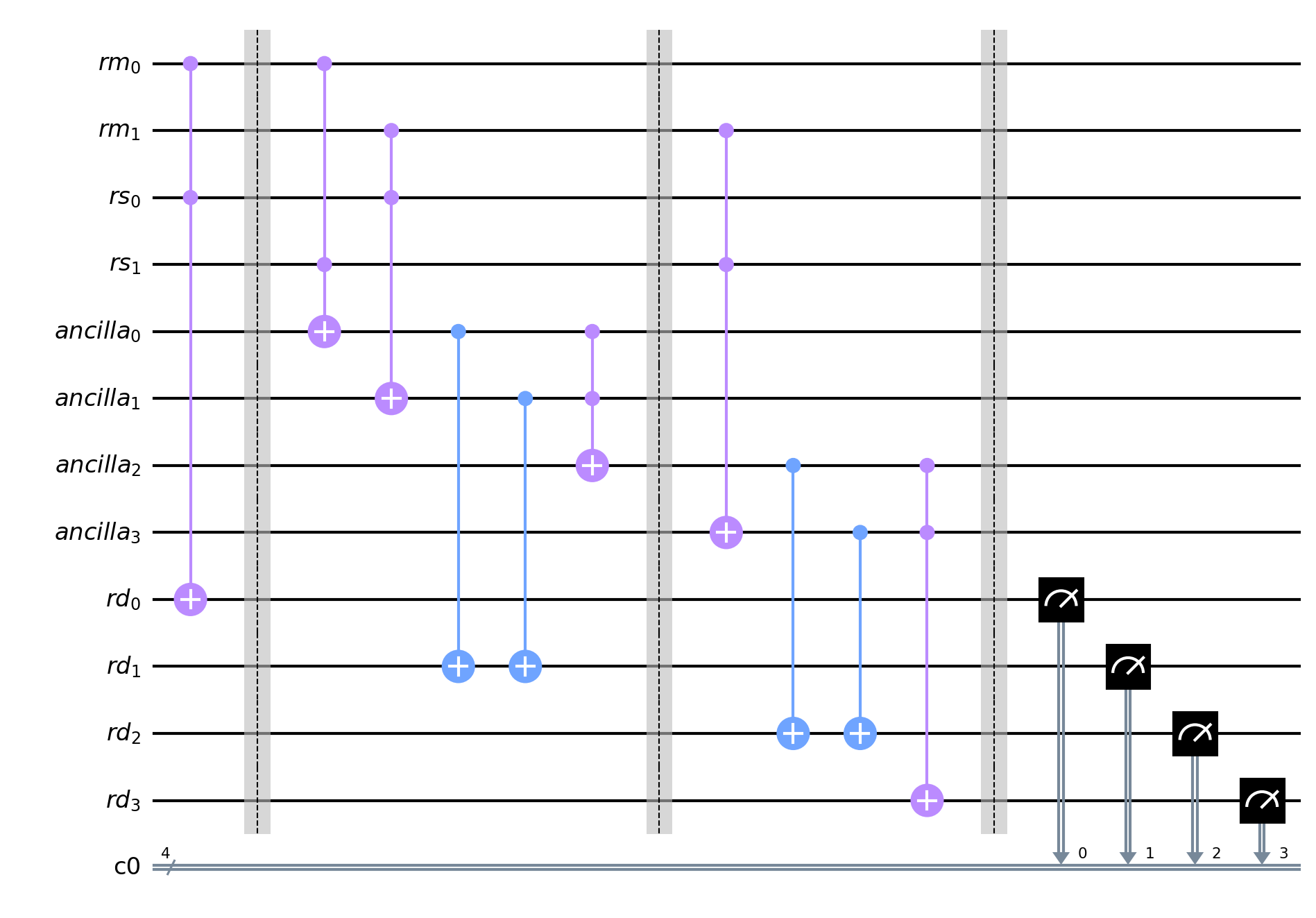}
    \caption{MUL Quantum Equivalent Circuit: The logic for multiplication is $R_0 \textrm{\&=} A_0 \land B_0, R_1 \textrm{\&=} (A_0 \land B_1) + (A_1 \land B_0), R_2 \textrm{\&=} (A_1 \land B_1) + \text{Carry from } R_1, R_3 \textrm{\&=} \text{Carry from } R_2$}
    \label{fig:MUL_quantum}
\end{figure*}

\textbf{MUL:} Figure \ref{fig:MUL_quantum} illustrates the mapping of the assembly MUL instruction to its quantum near-equivalent. The circuit uses two input registers, \textit{rm} and \textit{rs}, and employs Boolean operations to compute each bit of the multiplication efficiently according to the logic: $R_0 \textrm{\&=} A_0 \land B_0, R_1 \textrm{\&=} (A_0 \land B_1) + (A_1 \land B_0), R_2 \textrm{\&=} (A_1 \land B_1) + \text{Carry from } R_1, R_3 \textrm{\&=} \text{Carry from } R_2$ using CNOT and MCT.

\begin{figure}[htp]
    \centering
    \includegraphics[width=8cm]{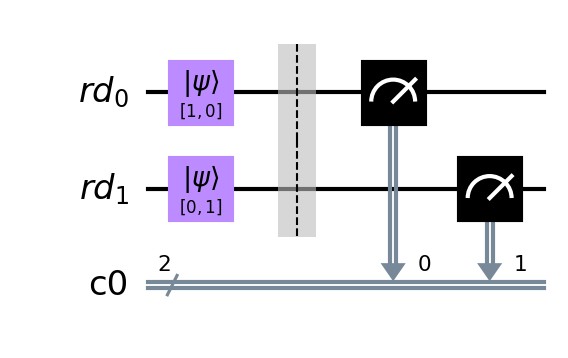}
    \caption{MOV Quantum Equivalent Circuit: Assigns values to Rd.}
    \label{fig:MOV_quantum}
\end{figure}

\textbf{MOV:} Figure \ref{fig:MOV_quantum} depicts the mapping of the assembly MOV instruction to its quantum near-equivalent. The circuit utilizes a single register, \textit{rd}, and applies an initialization gate to assign it a classically known value. If applied improperly, this operation may cause decoherence.

\begin{figure}[htp]
    \centering
    \includegraphics[width=8cm]{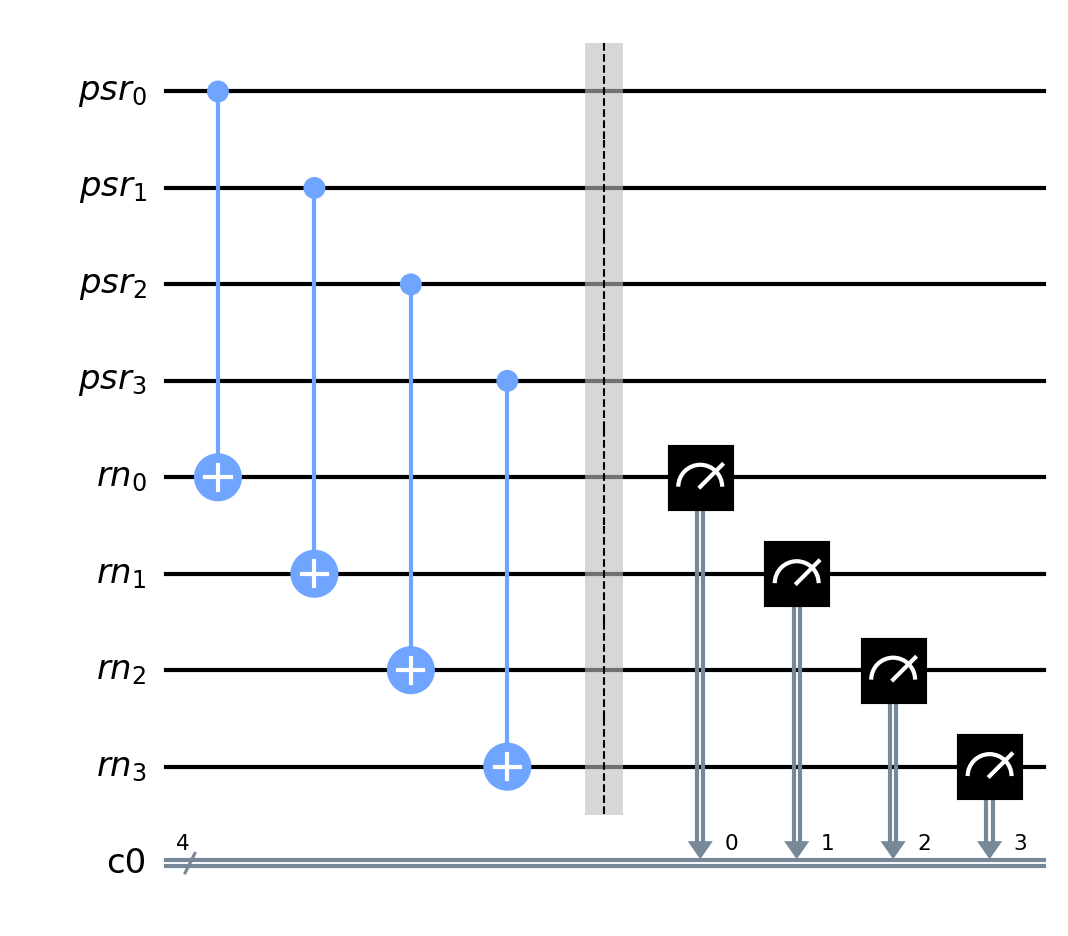}
    \caption{MRS Quantum Equivalent Circuit: Moves the value of the PSR register to the Rn register using controlled-not gates.}
    \label{fig:MRS_quantum}
\end{figure}

\textbf{MRS:} Figure \ref{fig:MRS_quantum} illustrates the mapping of the assembly MRS instruction to its quantum near-equivalent. The circuit uses a single input register, \textit{rn}, and transfers the bits from the PSR register to \textit{rn} using CNOT gates.

\begin{figure}[htp]
    \centering
    \includegraphics[width=8cm]{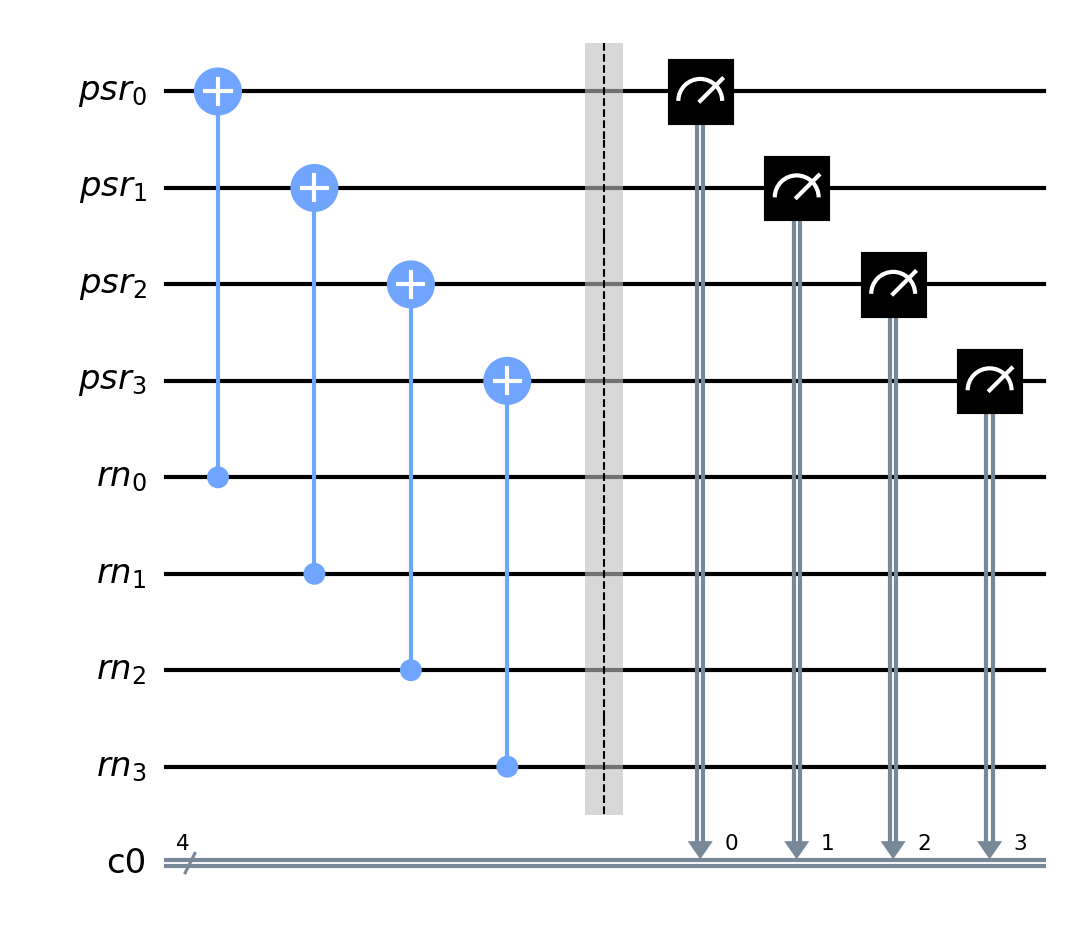}
    \caption{MSR Quantum Equivalent Circuit: Moves the value of the Rn register to the PSR register using controlled-not gates.}
    \label{fig:MSR_quantum}
\end{figure}

\textbf{MSR:} Figure \ref{fig:MSR_quantum} depicts the mapping of the assembly MSR instruction to its quantum near-equivalent. The circuit uses a single input register, \textit{rn}, and transfers its bits to the PSR register via CNOT gates.

\begin{figure}[htp]
    \centering
    \includegraphics[width=8cm]{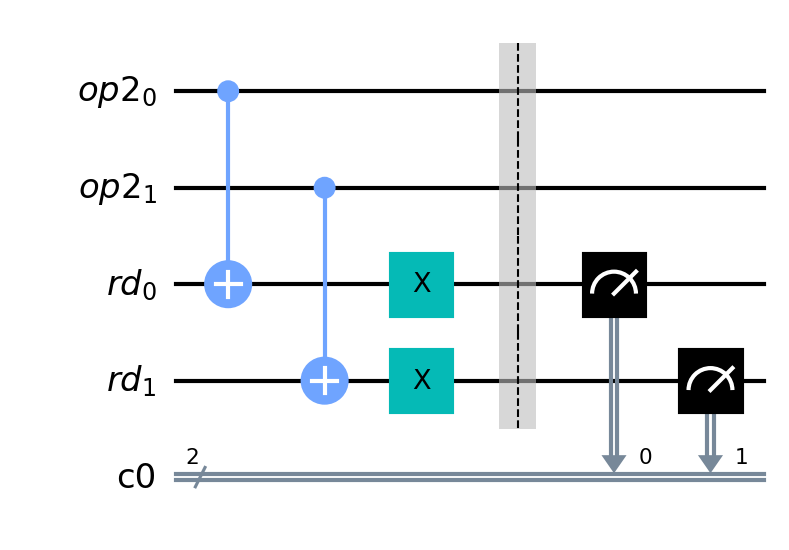}
    \caption{MVN Quantum Equivalent Circuit: Moves the inverted value from Op2 to Rd using controlled-not and X gates.}
    \label{fig:MVN_quantum}
\end{figure}

\textbf{MVN:} Figure \ref{fig:MVN_quantum} illustrates the mapping of the assembly MVN instruction to its quantum near-equivalent. The circuit uses a single register, \textit{rd}, and applies an initialization gate to assign it the classically known value in inverted form. If applied improperly, this operation may induce decoherence.

\begin{figure}[htp]
    \centering
    \includegraphics[width=8cm]{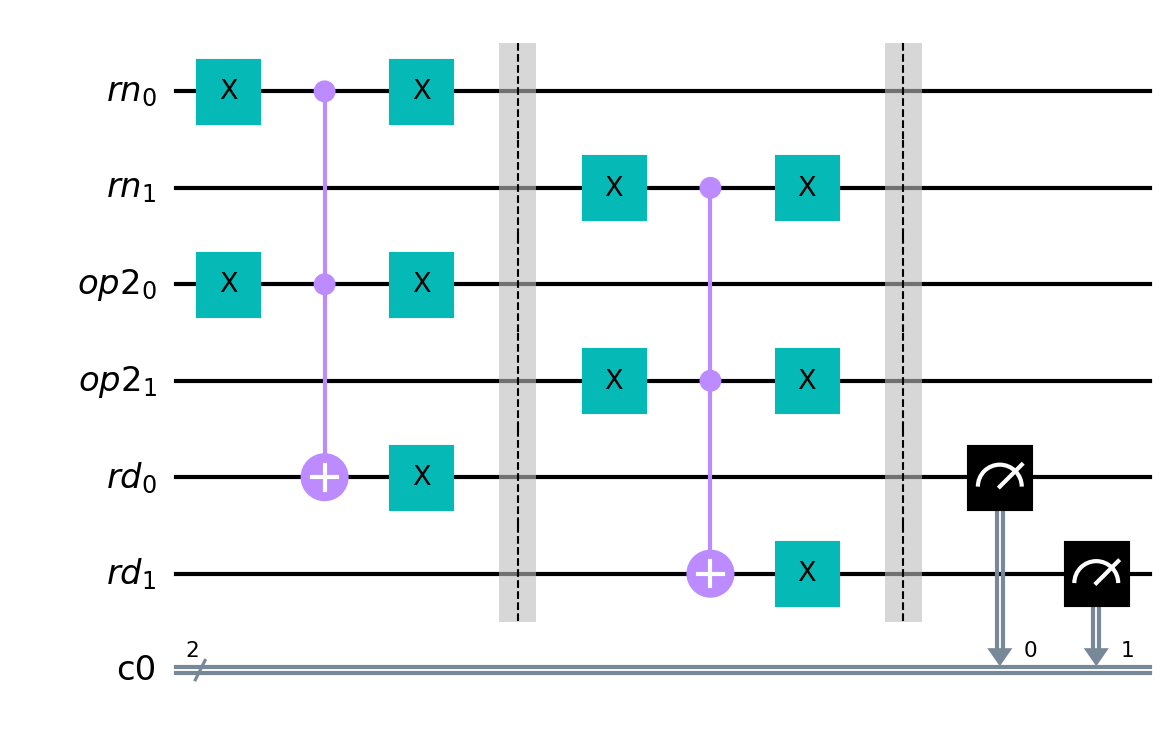}
    \caption{ORR Quantum Equivalent Circuit: Bitwise $Rd = Rn \vee Op2$ using X and Multiply-Controlled Toffoli gates.}
    \label{fig:ORR_quantum}
\end{figure}

\textbf{ORR:} Figure \ref{fig:ORR_quantum} depicts the mapping of the assembly ORR instruction to its quantum near-equivalent. The circuit utilizes two input registers, \textit{rn} and \textit{op2}, and implements the bitwise OR operation using a combination of X and MCT gates.

\begin{figure}[htp]
    \centering
    \includegraphics[width=8cm]{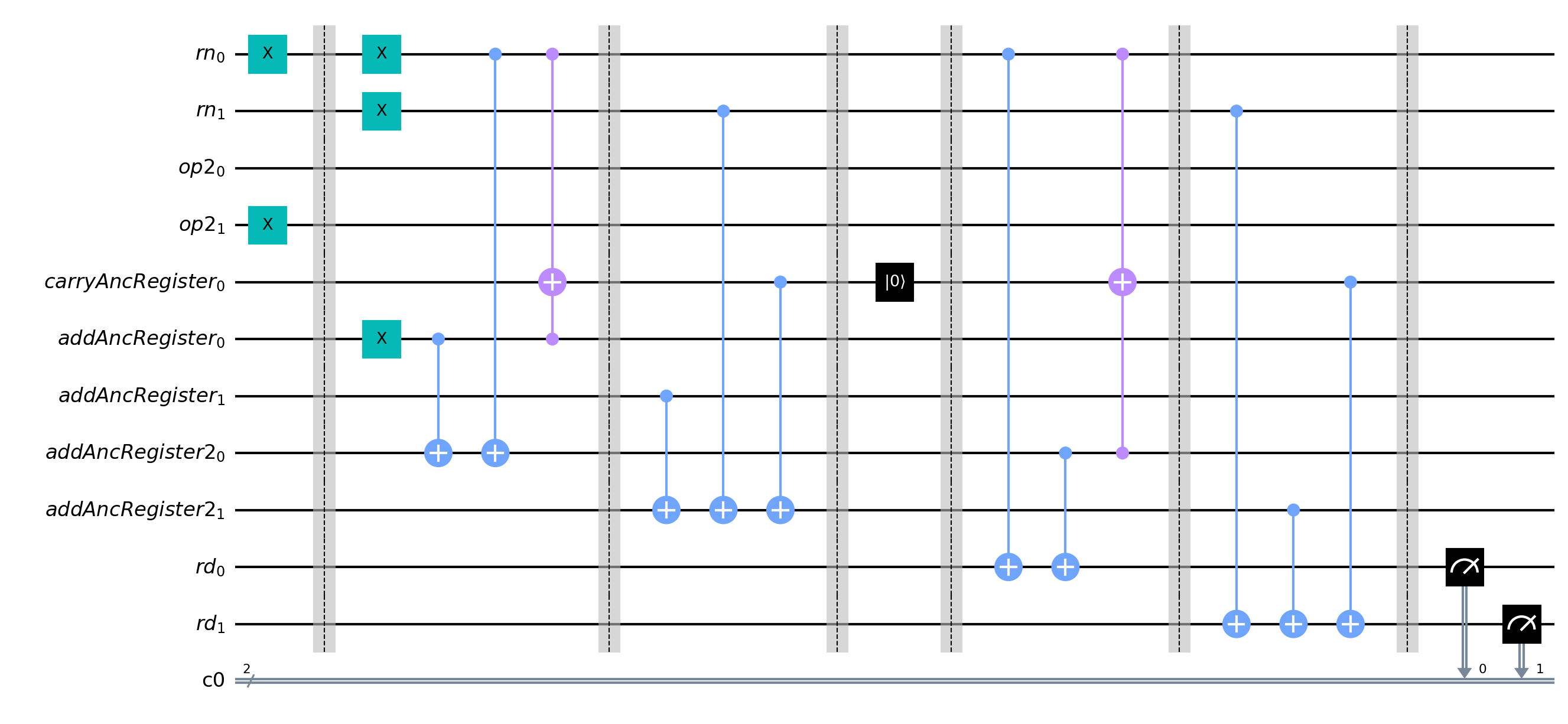}
    \caption{RSB Quantum Equivalent Circuit: $Rd = Op2 - Rn$ using X, controlled-not, multiply-controlled Toffoli, and reset gates.}
    \label{fig:RSB_quantum}
\end{figure}

\textbf{RSB:} Figure \ref{fig:RSB_quantum} illustrates the mapping of the assembly RSB instruction to its quantum near-equivalent. The circuit includes two input registers, \textit{rn} and \textit{op2}, and performs the operation by negating the value in \textit{rn} and then adding it to \textit{op2} using a combination of X, CNOT, and MCT gates.

\begin{figure*}[htp]
    \centering
    \includegraphics[width=18cm]{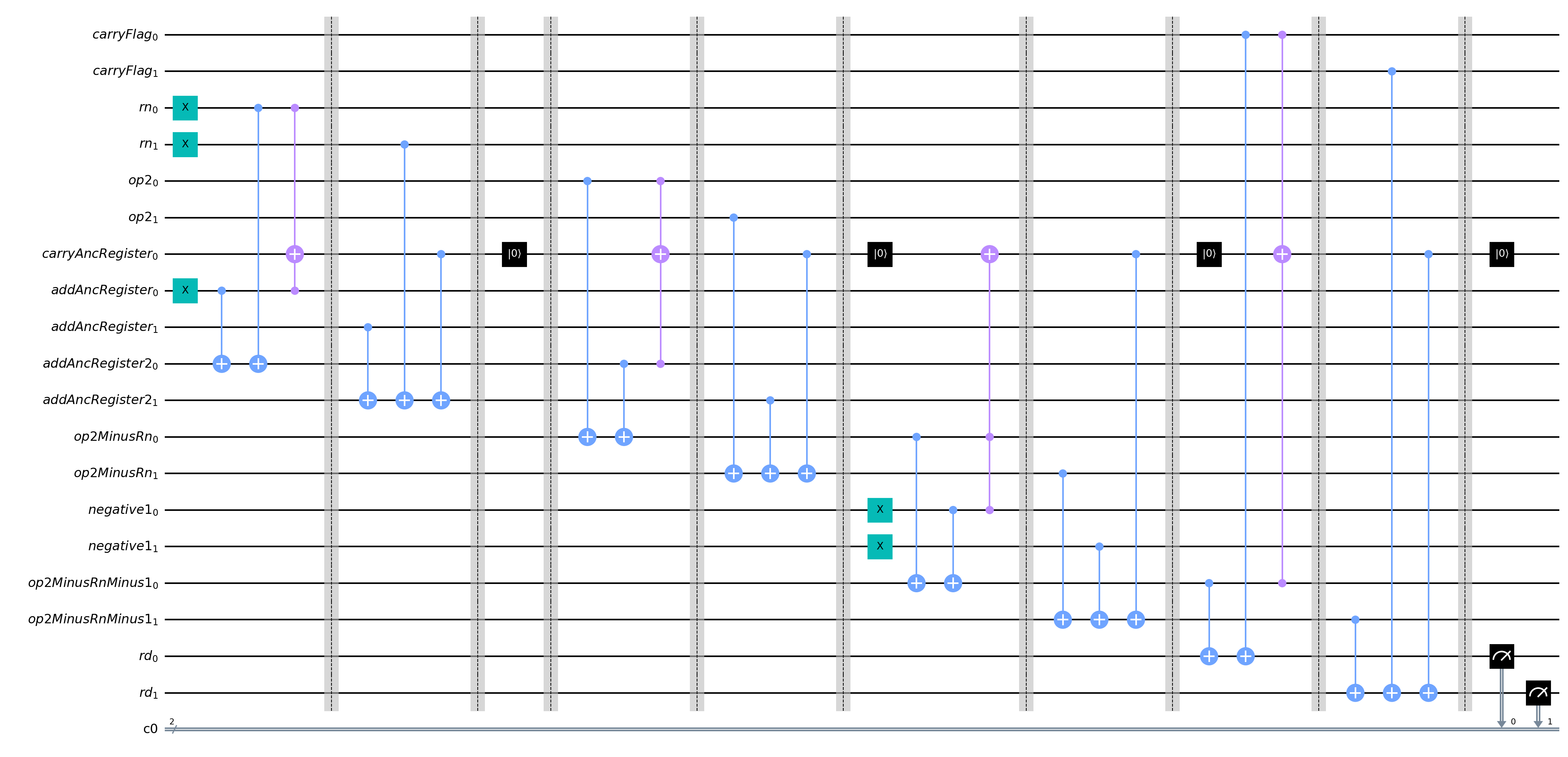}
    \caption{RSC Quantum Equivalent Circuit: $Rd := Op2 - Rn + Carry - 1$ using X, controlled-not, multiply-controlled Toffoli, and reset gates.}
    \label{fig:RSC_quantum}
\end{figure*}

\textbf{RSC:} Figure \ref{fig:RSC_quantum} illustrates the mapping of the assembly RSC instruction to its quantum near-equivalent. The circuit utilizes two input registers, \textit{rn} and \textit{op2}, and implements the operation by negating \textit{rn}, adding it to \textit{op2}, and then incorporating the carry flag minus one, using a combination of X, CNOT, and MCT gates.

\begin{figure*}[htp]
    \centering
    \includegraphics[width=18cm]{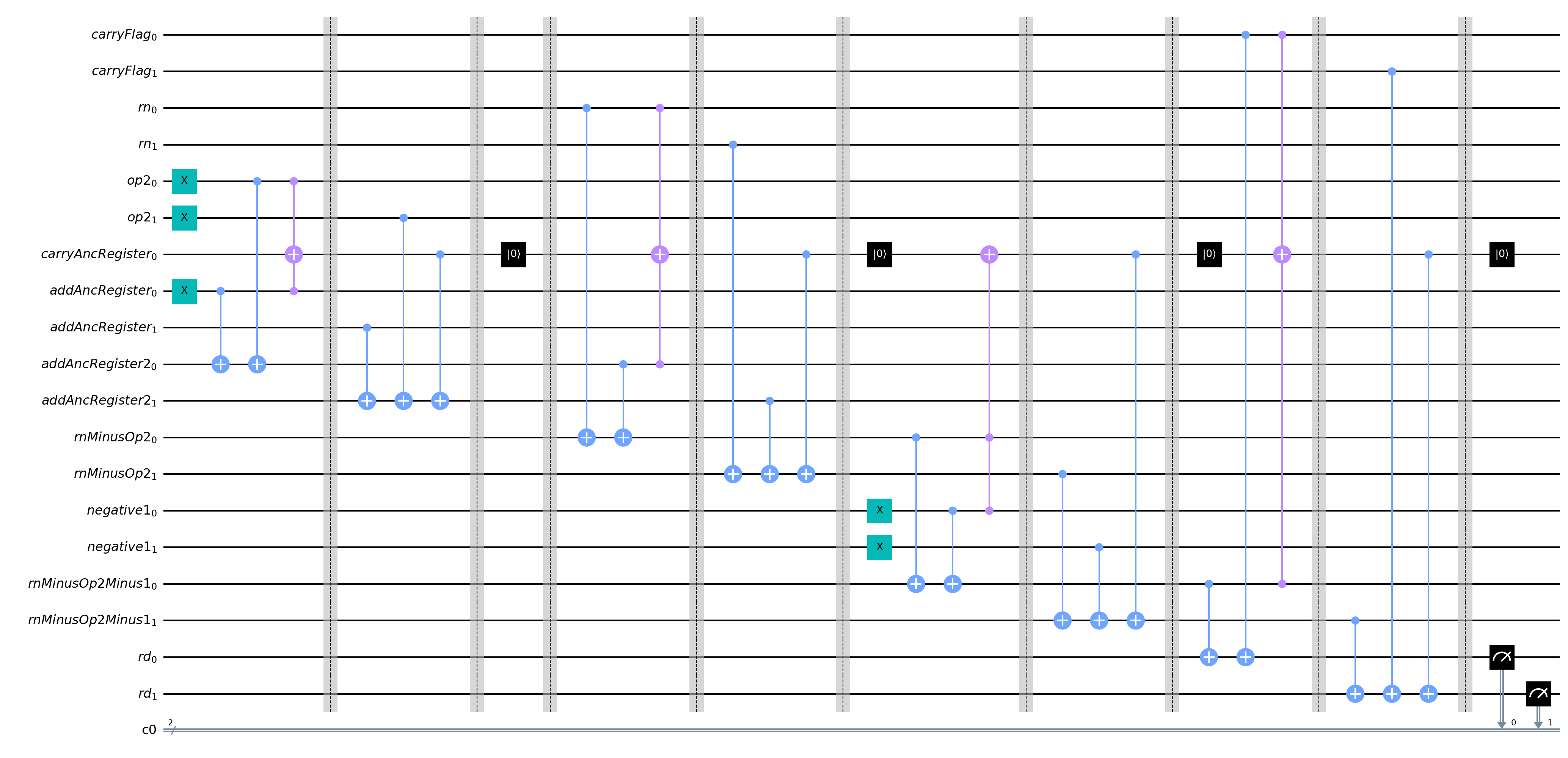}
    \caption{SBC Quantum Equivalent Circuit: $Rd := Rn - Op2 - 1 - Carry$ using X, controlled-not, multiply-controlled Toffoli, and reset gates.}
    \label{fig:SBC_quantum}
\end{figure*}

\textbf{SBC:} Figure \ref{fig:SBC_quantum} depicts the mapping of the assembly SBC instruction to its quantum near-equivalent. The circuit uses two input registers, \textit{rn} and \textit{op2}, and performs the operation by negating \textit{op2}, adding it to \textit{rn}, and then incorporating the carry flag minus one, using a combination of X, CNOT, and MCT gates.

\begin{figure}[htp]
    \centering
    \includegraphics[width=8cm]{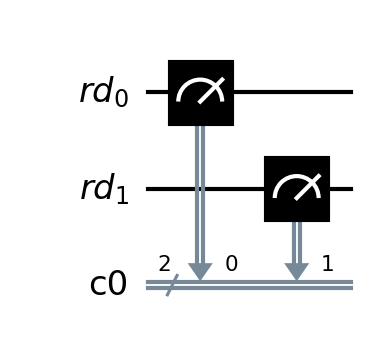}
    \caption{STR Quantum Equivalent Circuit: Classically stores a quantum register.}
    \label{fig:STR_quantum}
\end{figure}

\textbf{STR:} Figure \ref{fig:STR_quantum} illustrates the mapping of the assembly STR instruction to its quantum near-equivalent. The circuit uses a single input register, \textit{rd}, and applies a measurement gate to store its value classically. This operation inherently causes decoherence.

\begin{figure*}[htp]
    \centering
    \includegraphics[width=18cm]{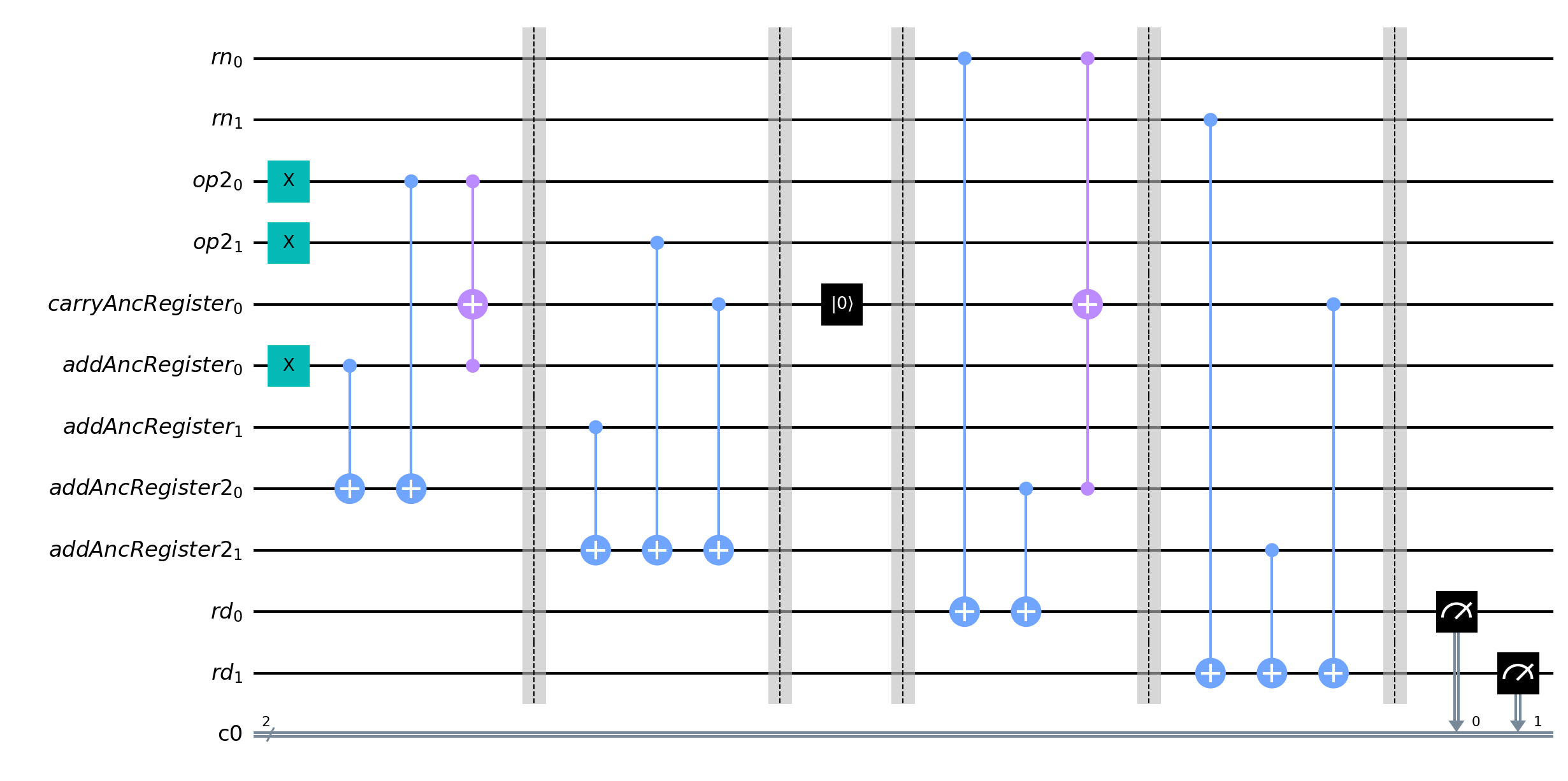}
    \caption{SUB Quantum Equivalent Circuit: $Rd := Rn - Op2$ using X, controlled-not, multiply-controlled Toffoli, and reset gates.}
    \label{fig:SUB_quantum}
\end{figure*}

\textbf{SUB:} Figure \ref{fig:SUB_quantum} depicts the mapping of the assembly SUB instruction to its quantum near-equivalent. The circuit uses two input registers, \textit{rn} and \textit{op2}, and performs the subtraction by negating \textit{op2} and adding it to \textit{rn} using a combination of X, CNOT, and MCT gates.

\begin{figure*}[htp]
    \centering
    \includegraphics[width=18cm]{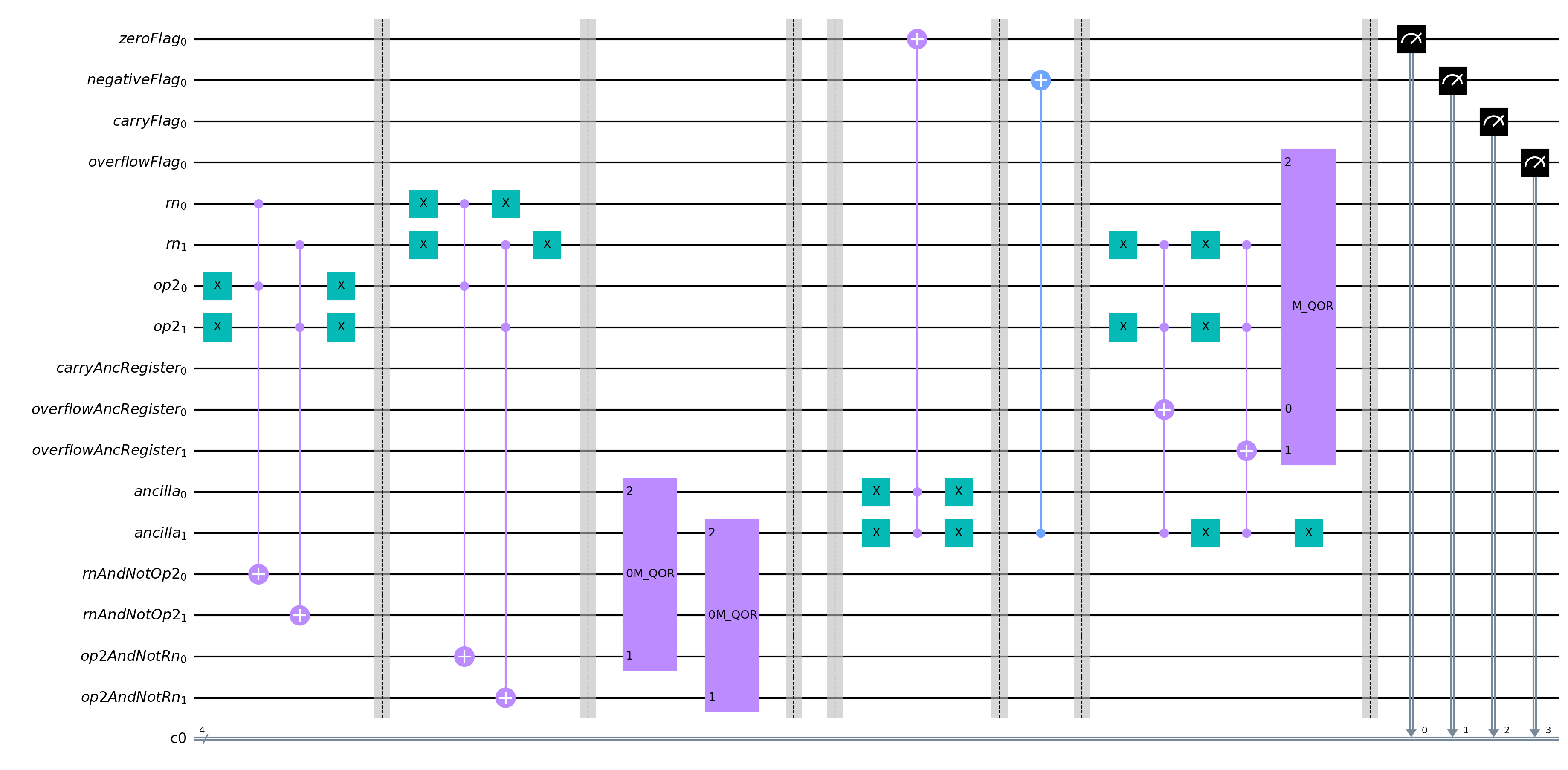}
    \caption{TEQ Quantum Equivalent Circuit: Performs a bitwise exclusive OR on Rn and Op2, then assigns the Zero, Negative, Carry, and Overflow flags accordingly using X, controlled-not, and MCT gates.}
    \label{fig:TEQ_quantum}
\end{figure*}

\textbf{TEQ:} Figure \ref{fig:TEQ_quantum} illustrates the mapping of the assembly TEQ instruction to its quantum near-equivalent. The circuit uses two input registers, \textit{rn} and \textit{op2}, and performs a bitwise exclusive OR between them. The \textit{Carry} flag is set if the operation produces a carry, the \textit{Zero} flag is determined by checking whether all result bits are zero, the \textit{Negative} flag is set based on the most significant bit (MSB), and the \textit{Overflow} flag is computed by XORing the MSBs of the input registers.

\begin{figure*}[htp]
    \centering
    \includegraphics[width=18cm]{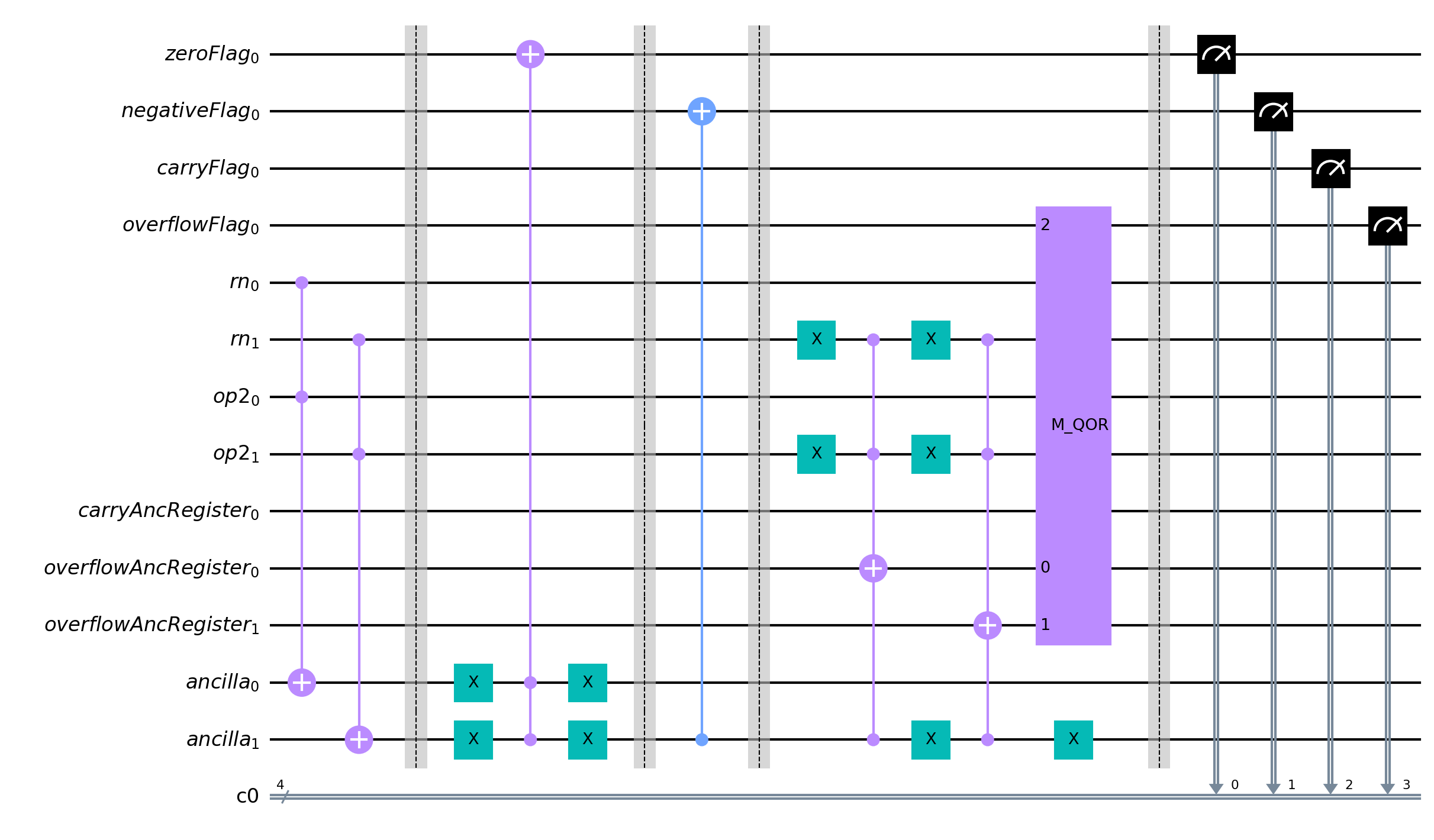}
    \caption{TST Quantum Equivalent Circuit: Performs a bitwise AND on Rn and Op2, then assigns the Zero, Negative, Carry, and Overflow flags accordingly using X, controlled-not, and MCT gates.}
    \label{fig:TST_quantum}
\end{figure*}

\textbf{TST:} Figure \ref{fig:TST_quantum} illustrates the mapping of the assembly TST instruction to its quantum near-equivalent. The circuit uses two input registers, \textit{rn} and \textit{op2}, and performs a bitwise AND between them. The \textit{Carry} flag is set if the operation generates a carry, the \textit{Zero} flag is determined by checking whether all result bits are zero, the \textit{Negative} flag is based on the most significant bit (MSB), and the \textit{Overflow} flag is computed by XORing the MSBs of the input registers.

\section{Evaluation: Calculating the Fibonacci Sequence}
\label{evaluation}

With a significant number of assembly instructions now mapped to their quantum equivalents, we can demonstrate their practical utility. One non-trivial example is computing the Fibonacci sequence. The first five Fibonacci numbers are 0, 1, 1, 2, 3 \cite{fibonacci}. The following presents an assembly program for calculating the first $N$ Fibonacci numbers.

\begin{lstlisting}[language=C]
    MOV R1, #0      ; Initialize R1 with 0, F(0)
    MOV R2, #1      ; Initialize R2 with 1, F(1)
    MOV R3, #N      ; R3 holds the value of N, the number of Fibonacci numbers to generate
    MOV R4, #1      ; Initialize loop counter R4 with 1

FIB_LOOP:
    CMP R4, R3      ; Compare loop counter with N
    BGE FIB_DONE    ; If counter >= N, finish

    ADD R5, R1, R2  ; R5 = R1 + R2, calculate the next Fibonacci number
    MOV R1, R2      ; Update R1 to the next Fibonacci number
    MOV R2, R5      ; Update R2 to the new next Fibonacci number

    ADD R4, R4, #1  ; Increment the counter
    B FIB_LOOP      ; Repeat the loop

FIB_DONE:
    ; At this point, R1 contains the last Fibonacci number generated
    ; The program ends here

    NOP             ; No Operation (End of program)
\end{lstlisting}




Next, we transform the program into a form that can be directly mapped to a quantum circuit. Several modifications are required:

\begin{enumerate}
\item Unravel the loop, eliminating the counter, labels, and branch instructions.
\item Remove all NOP instructions.
\item Due to qubit limitations, compute only up to the fifth Fibonacci number.
\end{enumerate}

The assembly program below incorporates these adjustments.

\begin{lstlisting}[language=C]
    MOV R1, #0      ; Initialize R1 with 0, F(0)
    MOV R2, #1      ; Initialize R2 with 1, F(1)

    ; calculate F(2)
    ADD R3, R1, R2  ; R3 = R1 + R2, calculate the next Fibonacci number
    MOV R1, R2      ; Update R1 to the next Fibonacci number
    MOV R2, R3      ; Update R2 to the new next Fibonacci number

    ; calculate F(3)
    ADD R3, R1, R2  ; R3 = R1 + R2, calculate the next Fibonacci number
    MOV R1, R2      ; Update R1 to the next Fibonacci number
    MOV R2, R3      ; Update R2 to the new next Fibonacci number

    ; calculate F(4)
    ADD R3, R1, R2  ; R3 = R1 + R2, calculate the next Fibonacci number
    MOV R1, R2      ; Update R1 to the next Fibonacci number
    MOV R2, R3      ; Update R2 to the new next Fibonacci number
\end{lstlisting}

\begin{figure*}[htp]
    \centering
    \includegraphics[width=18cm]{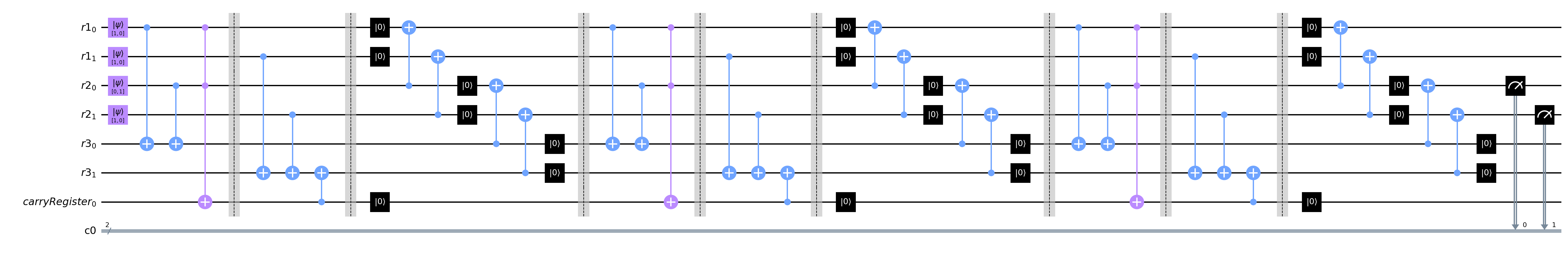}
    \caption{Fibonacci Coherence Breaking Quantum Equivalent Circuit}
    \label{fig:fibonacci_coherence_breaking}
\end{figure*}

\begin{figure}[htp]
    \centering
    \includegraphics[width=8cm]{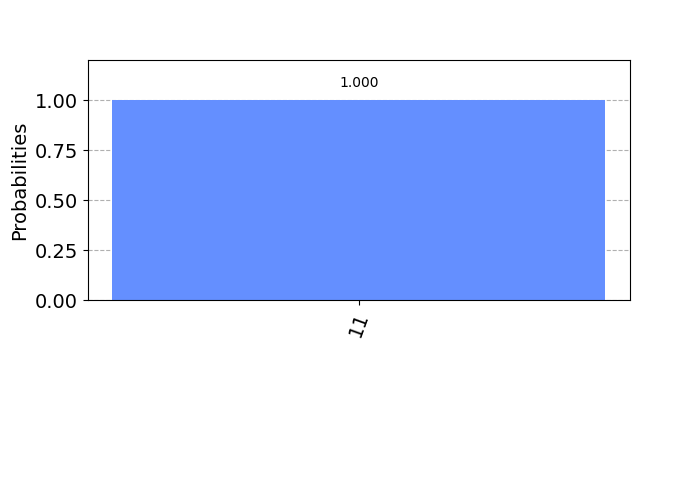}
    \caption{Fibonacci Coherence Breaking Quantum Equivalent Circuit Simulation Results}
    \label{hist:fibonacci_coherence_breaking}
\end{figure}

The corresponding quantum circuit is presented in Figure \ref{fig:fibonacci_coherence_breaking}, with its simulation results displayed in Figure \ref{hist:fibonacci_coherence_breaking}. The results confirm that the quantum circuit accurately computes the fifth Fibonacci number (3) using a coherence-breaking approach.

Reset gates disrupt coherence, which may be undesirable. To preserve coherence while calculating the Fibonacci sequence, the assembly program can be modified to avoid using reset gates and to ensure that qubits are not reused during computation. These modifications are incorporated in the following program:

\begin{lstlisting}[language=C]
MOV R1, #0       ; Initialize R1 with 0, F(0)
MOV R2, #1       ; Initialize R2 with 1, F(1)

; calculate F(2)
ADD R3, R1, R2   ; R3 = R1 + R2, calculate the next Fibonacci number (F(2))

; calculate F(3)
ADD R4, R2, R3   ; R4 = R2 + R3, calculate the next Fibonacci number (F(3))

; calculate F(4)
ADD R5, R3, R4   ; R5 = R3 + R4, calculate the next Fibonacci number (F(4))
\end{lstlisting}

\begin{figure*}[htp]
    \centering
    \includegraphics[width=18cm]{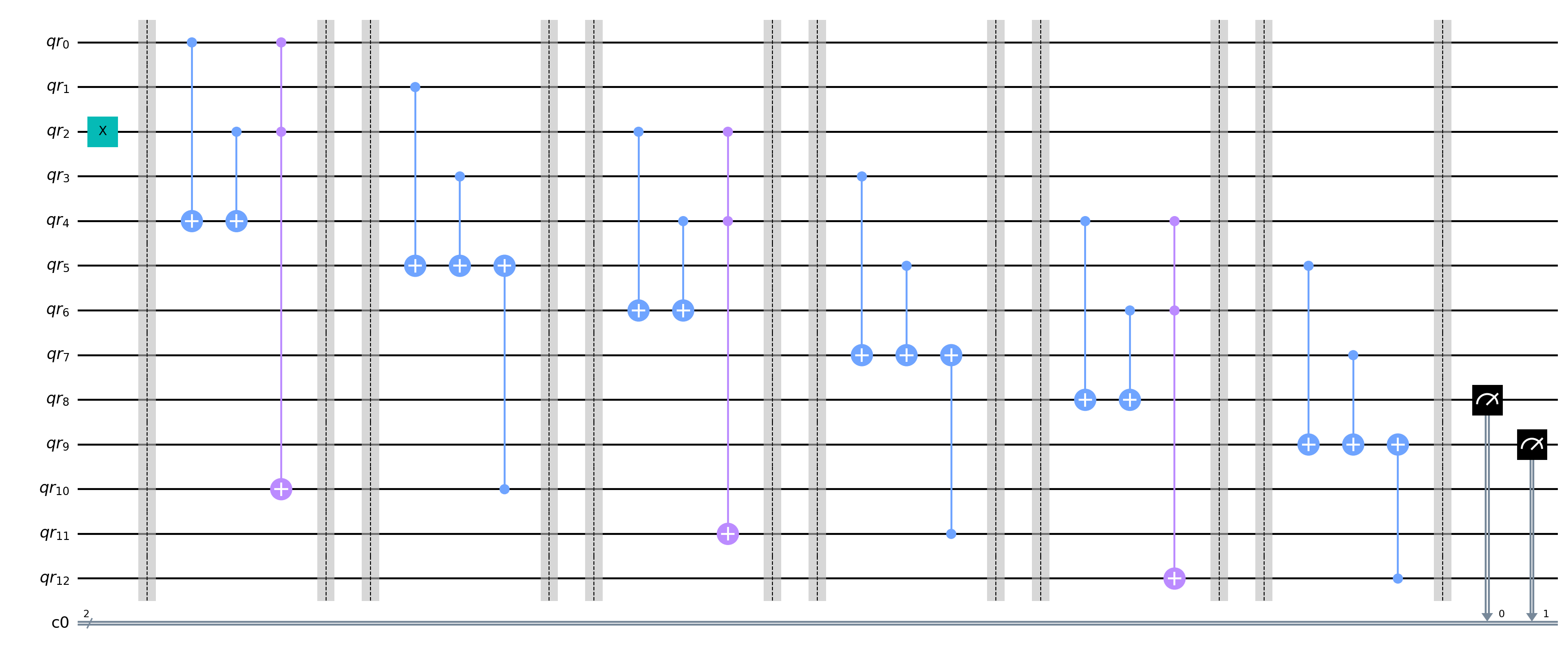}
    \caption{Fibonacci Coherence Maintaining Quantum Equivalent Circuit}
    \label{fig:fibonacci}
\end{figure*}

\begin{figure}[htp]
    \centering
    \includegraphics[width=8cm]{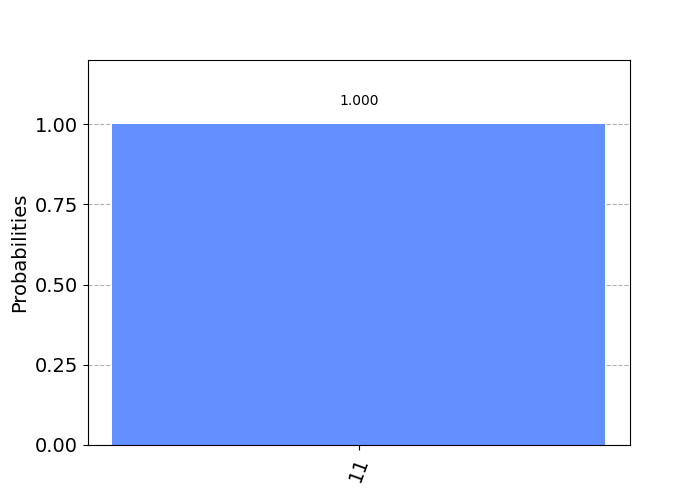}
    \caption{Fibonacci Coherence Maintaining Quantum Equivalent Circuit Simulation Results}
    \label{hist:fibonacci}
\end{figure}

Figure \ref{fig:fibonacci} shows the coherence maintaining circuit and Figure \ref{hist:fibonacci} shows the simulation results.

\clearpage

\section{Grover's Algorithm Implementation}
\label{grovers}

Although the mapping of assembly instructions to quantum operations is useful for constructing Grover’s Algorithm oracles, Grover’s Algorithm also requires additional operations. To address these needs, new quantum operations were implemented. Table \ref{quantum_specific} lists these added operations.


\begin{table*}[!htbp]
\centering
\caption{\label{TableQuantum} Quantum-Specific Operations}
\label{quantum_specific}
\small
\begin{tabular}{|lp{3cm}p{3cm}p{4cm}l|}
\hline
Mnemonic & Instruction & Action & Example & Coherence Maintaining? \\
\hline
HAD & Apply Hadamard gates & Rd := $\ket{+}$ & HAD Rd & Yes \\
XXX & Apply Pauli X gates & Rd := qc.x(Rd) & XXX Rd & Yes \\
TGT & Apply CNot gate from flag or Rd[0] to target & qc.cx(Rd[0], target) & TGT Rd & Yes \\
DIF & Apply Diffuser & Diffuser applied to {RList} & DIF {RList} & Yes \\
BAR & N/A & Create barrier in circuit diagram & BAR & N/A \\
\hline
\end{tabular}
\end{table*}


Additionally, the oracle begins with the instruction ``ORACLE," ends with ``END\_ORACLE," and the reverse of the first half of the oracle is applied using ``REVERSE\_ORACLE."

\begin{figure}[htp]
    \centering
    \includegraphics[width=8cm]{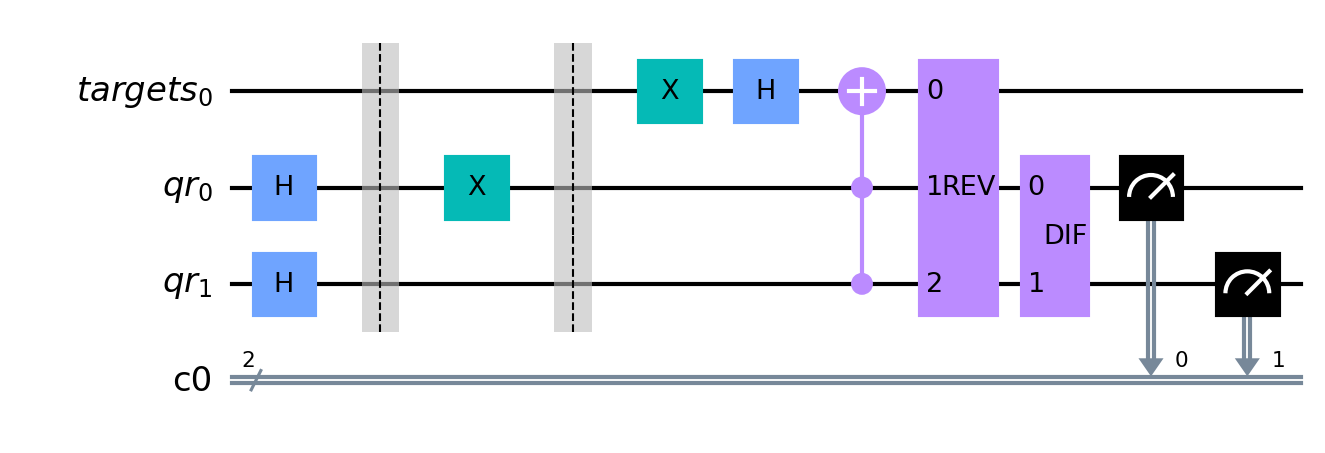}
    \caption{Simple Grover's Algorithm Circuit}
    \label{fig:grovers_simple}
\end{figure}

\begin{figure}[htp]
    \centering
    \includegraphics[width=8cm]{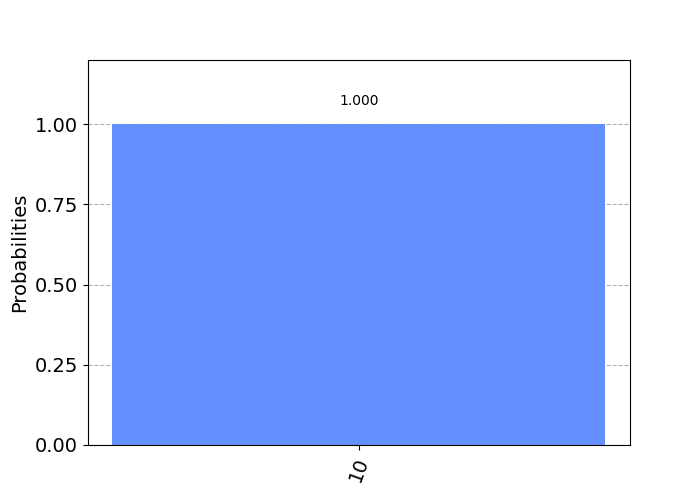}
    \caption{Simple Grover's Algorithm Circuit Simulation Results}
    \label{hist:grovers_simple}
\end{figure}

These operations can now be used to implement Grover’s Algorithm. The following code generates the circuit shown in Figure \ref{fig:grovers_simple}, with its simulation results displayed in Figure \ref{hist:grovers_simple}. The results demonstrate that Grover’s Algorithm can be executed successfully using only these assembly-to-quantum transformations.

\begin{lstlisting}[language=C]
{"register_size": 2}
;hadamard
HAD R1
BAR
ORACLE
	MOV R1, #1
END_ORACLE
BAR
;applies all bits from R1 to the target using a Multiply-Controlled Toffoli Gate
MCT R1
REVERSE_ORACLE
;diffuser
DIF {R1}
STR CR1, R1
\end{lstlisting}

\section{Open Source Project}
\label{open_source}

This compiler is open-source and available under the MIT License at \\https://github.com/arhaverly/AssemblyToQuantumCompiler.

The compiler takes a simplified ARM assembly file as input and generates the corresponding quantum circuit. Parameters can be configured via a JSON block at the beginning of the file, with currently supported options including register size, decoding, execution, and display settings. Example files are included in the repository.

\section{Conclusion}
\label{conclusion}
This study achieved the transformation of ARM assembly instructions into a quantum computing format, focusing on their application in quantum algorithms. The main accomplishment was mapping ARM instructions to quantum operations. The Fibonacci sequence computation, simulated using these quantum mappings, served as a practical demonstration of this approach. This simulation not only proved the technical feasibility but also highlighted the potential for applying classical programming techniques in quantum computing. Additionally, Grover's Algorithm was implemented using quantum-specific commands. The results are particularly relevant for simplifying algorithms like Grover's in quantum environments. These transformations were implemented to make an open-source assembly to quantum compiler: github.com/arhaverly/AssemblyToQuantumCompiler. This research contributes to integrating classical and quantum computing, paving the way for further explorations in this area.

\section{Acknowledgements}
Portions of this work were previously included in the first author's doctoral dissertation at Mississippi State University, 2025.

\begin{spacing}{0.9}

\bibliographystyle{IEEEbib}
\bibliography{strings}


\end{spacing}
\end{document}